\documentclass[a4paper,11pt]{article}
\usepackage{pos}
\usepackage{mathtools}
\usepackage{physics}
\usepackage{tensor}
\usepackage{faktor}
\usepackage{cleveref}
\usepackage{amsfonts}
\usepackage{amsmath}

\newcommand{\D}{\text{d}}
\newcommand{\beq}{\begin{equation}}
\newcommand{\eeq}{\end{equation}}
\newcommand{\beqn}{\begin{eqnarray}}
\newcommand{\eeqn}{\end{eqnarray}}

\newcommand{\cL}{{\cal L}}
\newcommand{\fL}{\mathfrak{L}}
\newcommand{\he}{ \ \hat{=} \ }

\newcommand{\widesim}[2][1.5]{
  \mathrel{\overset{#2}{\scalebox{#1}[1]{$\sim$}}}
}

\usepackage{pict2e}

\makeatletter
\DeclareRobustCommand{\loplus}{\mathbin{\mathpalette\dog@lsemi{+}}}

\newcommand{\dog@lsemi}[2]{\dog@semi{#1}{#2}{270,90}}
\newcommand{\dog@semi}[3]{%
  \begingroup
  \sbox\z@{$\m@th#1#2$}%
  \setlength{\unitlength}{\dimexpr\ht\z@+\dp\z@\relax}%
  \makebox[\wd\z@]{\raisebox{-\dp\z@}{%
    \begin{picture}(1,1)
    \linethickness{\variable@rule{#1}}
    \roundcap
    \put(0.5,0.5){\makebox(0,0){\raisebox{\dp\z@}{$\m@th#1#2$}}}
    \put(0.5,0.5){\arc[#3]{0.5}}
    \end{picture}%
  }}%
  \endgroup
}
\newcommand{\variable@rule}[1]{%
  \fontdimen8  
  \ifx#1\displaystyle\textfont3\else
    \ifx#1\textstyle\textfont3\else
      \ifx#1\scriptstyle\scriptfont3\else
        \scriptscriptfont3\relax
  \fi\fi\fi
}
\makeatother

\newcommand{\defeq}{\vcentcolon=}

\hypersetup{
colorlinks=true,
linkcolor=blue,
citecolor=red,
urlcolor=cyan
}

\newcommand{\eom}{\text{e.o.m.}}

\newcommand{\ldv}[1]{\mathcal{L}_{#1}}
\newcommand{\intpull}[2]{\int_{#1} \phi^*}

\title{Cornering Quantum Gravity}

\author*[a]{Luca Ciambelli}
\author[b]{Alessandra D'Alise}
\author[b]{Vittorio D'Esposito}
\author[c]{Dušan Đorđević}
\author[d]{Diego Fern\'andez-Silvestre}
\author[e]{Ludovic Varrin}

\affiliation[a]{Perimeter Institute for Theoretical Physics,\\
31 Caroline St. N., Waterloo ON, Canada, N2L 2Y5}
\affiliation[b]{Dipartimento di Fisica “E.Pancini”, Universit\`a di Napoli Federico II, I-80125 Napoli, Italy\\ and\\ INFN, Sezione di Napoli, Complesso Universitario di Monte S. Angelo, Via Cintia Edificio 6, 80126 Napoli, Italy}

\affiliation[c]{Faculty of Physics, University of Belgrade, Studentski trg 12, Belgrade}

\affiliation[d]{Departamento de Matem\'aticas y Computaci\'on, Universidad de Burgos, 09001 Burgos, Spain}

\affiliation[e]{National Centre for Nuclear Research, Pasteura 7, 02-093 Warsaw, Poland}

\emailAdd{ciambelli.luca@gmail.com}

\abstract{
After introducing the covariant phase space calculus, Noether's theorems are discussed, with particular emphasis on Noether's second theorem and the role of gauge symmetries. This is followed by the enunciation of the theory of asymptotic symmetries, and later its application to gravity. Specifically, we review how the BMS group arises as the asymptotic symmetry group of gravity at null infinity. Symmetries are so powerful and constraining that memory effects and soft theorems can be derived from them. The lectures end with more recent developments in the field: the corner proposal as a unified paradigm for symmetries in gravity, the extended phase space as a resolution to the problem of charge integrability, and eventually the implications  of the corner proposal on quantum gravity. }

\FullConference{%
  Winter School of Theoretical Physics and Third Training School of COST Action CA18108\\
12–21 February 2023\\
Pałac Wojanów (Jelenia Góra), Poland,\\}

\begin{document}
\maketitle
\newpage
\clearpage
\pagenumbering{arabic}

\paragraph{Technical Note:} Subsecs. \ref{Sec1.1}, \ref{Noetherstheoremsandchargealgebra}, \ref{Sec1.4}, \ref{Sec2.1}, and Section \ref{Sec3} follow closely reference \cite{Ciambelli:2022vot}. We refer to it for a detailed introduction on the topics exposed  and a complete list of references.

\section*{Introduction}
The aim of these lectures is to provide students an introduction to the idea behind the \textit{Corner Proposal} for Quantum Gravity. Basically, the proposal is to gain physical intuition about the quantum nature of gravity by deriving a universal class of Noether charges, and their associated algebras, describing gravity on the classical level. Then, upon quantization, these are the observables that any theory of quantum gravity should encompass. Specifically, it is shown that the so-called \textit{Universal Corner Symmetry algebra} is maximal, namely it is independent of a particular setup. Therefore, any specification of a theory or a choice of a quantization scheme will give rise to a subset of this charge algebra.

The notions of symmetry and associated charges are therefore the building blocks of this proposal. These topics are discussed in \cref{Symmetries sec} within the formalism of the covariant phase space, introduced in \cref{Sec1.1}, which is more suitable for the task at hand. The characterization of Noether charges in this formalism is then discussed in \cref{Noetherstheoremsandchargealgebra}. To better understand how this formalism works, an application to gauge theories is presented in \cref{Sec1.3}. Then, the notion of asymptotic symmetries, mostly applied in the subsequent sections, is introduced in \cref{Sec1.4}.

In \cref{gravity sec} the application of the formalism introduced in \cref{Symmetries sec} to the gravitational setting is considered. The diffeomorphim invariance characterizing gravity is the source of some issues that are analyzed and addressed in \cref{Sec2.1}. Two physical results are then presented: in \cref{bms4}, the Bondi-van der Burg-Metzner-Sachs (BMS) algebra is introduced and it is then used to derive the memory effect, characterizing the passage of gravitational waves, in \cref{memory sec}; in \cref{soft theorems sec}, Weinberg's soft theorems are derived from symmetry arguments.

The last part of the lectures, namely \cref{Sec3}, is devoted to the corner proposal. In \cref{universalalgebra} the Universal Corner Symmetry algebra is derived. Then, in \cref{extendedphasespace} the problem of the integrability of charges is addressed by introducing the notion of \textit{extended phase space}. Finally, the corner proposal is stated and discussed in \cref{corner proposal sec}.

\section{Symmetries}\label{Symmetries sec}
Symmetries are one of the most powerful tools in the description of physical theories. From symmetry arguments physical laws can be derived and the solution to complicated problems can be simplified by restricting the allowed solutions to particular classes obeying a given symmetry. Symmetries can also be used to constraint physical theories: for instance, the number of allowed Lagrangians in relativistic quantum field theory is heavily limited by the requirement of Lorentz invariance. In these notes, we will adopt a geometric approach to the description of symmetries based on the so-called \emph{covariant phase space} formalism.

The first naive distinction that can be found in textbooks is between \emph{local} and \emph{global} symmetries.

\begin{itemize}
    \item Local symmetries are described as transformations characterized by spacetime dependent parameters. They are typically thought of as pure redundancies without any physical content.
    \item Global symmetries are described as transformations being specified by constant parameters. These are typically regarded as physical transformations acting non-trivially on physical systems.
\end{itemize}
The above distinction is somewhat confusing in that there are some contexts, specifically in the presence of \emph{boundaries}, when local symmetries are actually physical. The spacetime dependence of transformation parameters should not be the discerning feature in the distinction between physical symmetries and non-physical ones. The latter should be rather guided by the presence (or absence) of non-vanishing associated charges. Therefore, the fundamental tools in this distinction are Noether's theorems \cite{Noether_1971}.

There are two main frameworks in the study of Noether's theorems and symmetries:\footnote{As already mentioned, we refer to \cite{Ciambelli:2022vot} for a complete list of references.} 

\begin{itemize}
    \item In the Hamiltonian formalism, see \cite{REGGE1974286, BENGURIA197761},
    the evolution of the system is described by trajectories in phase space. The positive aspects of this formalism are that it is a very robust and unambiguous framework and, more importantly, it is the classical approach which is closest to quantum physics (it is indeed the one used within the canonical quantization procedure). However, there are also some important drawbacks, the most important of which for our purposes is that it breaks the manifest spacetime covariance. This is the main reason why we will not adopt this formalism.
    \item In the Lagrangian approach \cite{Noether_1971}, see also \cite{Harlow:2019yfa}, the phase space is in one to one correspondence with the set of solutions to the equations of motion. This is why this approach is manifestly covariant, indeed there is no specification of what time is but rather the physical system is described in terms of this abstract set of solutions. However, this comes with some negative aspects, mainly that this approach is somewhat more ambiguous and thus it is sometimes less intuitive.
\end{itemize}
As anticipated, we will adopt the Lagrangian formalism in these notes in that it allows us to apply the covariant phase space formalism which we will now introduce.

\subsection{Covariant phase space}\label{Sec1.1}

The idea of this formalism is to describe spacetime calculus (see for instance \cite{nakahara2003geometry}) and field space calculus (see \cite{Harlow:2019yfa} for a recent review) on the same footing.

\subsubsection{Spacetime (De Rham) calculus} 
Given a manifold $M$, it is possible to introduce the De Rham calculus on the space of forms. A $1$-form is a linear map from $TM$, the space of vector fields over $M$, to $C^\infty(M)$, the space of infinitely differentiable functions over $M$. The space of $1$-forms is dual to $TM$ and is therefore denoted by $T^*M$. A generic $p$-form is a linear, skew-symmetric map, from $p$ copies of $TM$ to $C^\infty(M)$. $p$ is also referred to as degree of the $p$-form. Given the skew-symmetric nature of these objects, the space of $p$-forms is denoted by $\Omega^p(M,\mathbb{R})=\bigwedge\nolimits^{p} T^*M$, where $\bigwedge$ denotes the skew-symmetric product. If $\text{dim}\, M = d$, it is clear that the maximum degree of a form is $d$. Such $d$-forms are called top forms. Therefore, we can build the space of all forms, called the \emph{De Rham complex}, as

\begin{equation}
    \Omega^\bullet (M,\mathbb{R})\defeq \bigoplus_{p=0}^{d}\bigwedge\nolimits^{p} T^*M \;,
\end{equation}
where $0$-forms are scalar functions, namely $\bigwedge\nolimits^{0} T^*M \defeq C^\infty(M)$. Observe that $\text{dim}\, \Omega^{d}(M,\mathbb{R}) = 1$, namely there exist only one top form, modulo function multiplication. On this complex, three derivations can be defined, namely the interior product, the exterior derivative, and the Lie derivative. 

The \textit{interior product} is defined as a contraction of a $p$-form with a vector field and it is a (anti-)derivation of degree $-1$, namely it decreases by one the degree of forms. If $X \in TM$, the interior product is a map

\begin{equation}
    \iota_X \vcentcolon \Omega^p(M,\mathbb{R}) \mapsto \Omega^{p-1}(M,\mathbb{R})
\end{equation}
defined as

\begin{equation}
    \iota_X \omega (Y_1,\ldots,Y_{p-1}) \defeq \omega (X,Y_1,\ldots,Y_{p-1})\;,
\end{equation}
where $\omega \in \Omega^{p}(M,\mathbb{R})$. If $\alpha$ is a $1$-form, then $\iota_X \alpha = \alpha (X)$, while $\iota_X f = 0 \;\, \forall\, f\in C^\infty(M)$. If $\alpha$ is a $p$-form and $\beta$ is a $q$-form, then the interior product satisfies the graded Leibnitz rule

\begin{equation}
    \iota_X(\alpha\wedge\beta) = \iota_X(\alpha)\wedge\beta +(-1)^p\alpha\wedge\iota_X\beta\;,
\end{equation}
from which it is clear that this map is an anti-derivation. Another important property of the interior product is that $\iota_X\iota_Y = -\iota_Y\iota_X$ from which it follows that $\iota_X \iota_X = 0$.

The \textit{exterior derivative} is an anti-derivation of degree $1$, namely it raises by one the degree of forms. It is a map

\begin{equation}\label{exterior derivative}
    \D \vcentcolon \Omega^p(M,\mathbb{R}) \mapsto \Omega^{p+1}(M,\mathbb{R})
\end{equation}
satisfying the graded Leibnitz rule

\begin{equation}
    \D (\alpha \wedge \beta) = \D\alpha\wedge\beta + (-1)^p \alpha\wedge\D\beta\;,
\end{equation}
where $\alpha$ and $\beta$ are a $p$-form and a $q$-form respectively. It is assumed that the exterior derivative is a coboundary, namely a nilpotent operator $\D^2 = 0$.

The third derivation that can be defined on the De Rham complex is the \textit{Lie derivative}, which is a derivation of degree $0$, it does not change the degree of forms. Given a vector field $X$, the Lie derivative along $X$ is given by the anti-commutator of the interior product and the exterior derivative

\begin{equation}\label{Lie der spacetime}
    \cL_X \defeq \D\iota_X +\iota_X\D \;. 
\end{equation}
This is typically referred to as \emph{Cartan's magic formula}. From this, we can also derive the full algebra satisfied by the three derivations we have defined, which turns out to be

\begin{equation}
    \comm{\cL_X}{\iota_X} = \comm{\cL_X}{\D} = 0 \;,\; \comm{\cL_X}{\iota_Y} = \comm{\iota_X}{\cL_Y}=\iota_{\comm{X}{Y}}\;.
\end{equation}
 All in all, the Lie derivative defines the directional derivative along the vector field $X$. Given a tensor $g$, for instance a metric defined on $M$, an \emph{isometry} of $g$ is a vector field $X$ such that 

\begin{equation}
    \cL_X g = 0\;,
\end{equation}
namely the tensor $g$ does not change along integral curves of $X$.

Until now we have only used intrinsic geometric definitions, without referring to any coordinate system. If $x^\mu$ are coordinates on $M$, then $\D x^\mu$ is a basis of $T^*M$ and $\partial_\mu$ is a basis of $TM$. Therefore, a generic $1$-form and a generic vector field are given by $\omega=\omega_\mu \D x^\mu$ and $X=X^\mu \partial_\mu$ respectively (implicit sum on repeated indexes). Therefore, the interior product yields

\begin{equation}
    \iota_X \omega = X^\mu \omega_\mu\;,
\end{equation}
i.e. the contraction. Analogously, the exterior derivative of a function yields

\begin{equation}
    \D f = \partial_\mu f \D x^\mu\;,
\end{equation}
which is nothing else than the differential of the function. If $g$ is a metric tensor defined on $M$, the top form, also called volume form, in coordinates is given by

\begin{equation}
    \sqrt{\abs{\det{g}}}\D x^1\wedge\ldots\wedge\D x^d = \sqrt{\abs{\det{g}}}\D^dx = \star 1\;,
\end{equation}
where $\star$ is the Hodge star which, for a generic $p$-form $\omega$, is defined as

\begin{equation}\label{star hodge}
    \star \vcentcolon \Omega^{p}(M,\mathbb{R})\mapsto \Omega^{d-p}(M,\mathbb{R}) \;\;\Big|\;\; \star \omega \defeq \frac{\sqrt{\abs{\det{g}}}}{(d-p)!}\omega_{\mu_1,\ldots,\mu_p}\tensor{\epsilon}{^{\mu_1,\ldots,\mu_p}}{_{\nu_1,\ldots,\nu_{d-p}}}\D x^{\nu_1}\wedge\ldots\wedge\D x^{\nu_{d-p}}\;,
\end{equation}
with $\epsilon$ being the Levi-Civita symbol.

To conclude this survey of spacetime calculus, we define the integration of forms. A $p$-form can be integrated over  $p$-dimensional sub-manifolds (i.e. on sub-manifolds on which it is a top form). For instance, a one form can be integrated over a curve $\gamma\vcentcolon[0,1]\mapsto M$ on the manifold $M$

\begin{equation}
    \int_\gamma \omega = \int_\gamma \omega_\mu \D x^\mu \Big|_\gamma = \int_{\gamma(0)=a}^{\gamma(1)=b} \omega_\mu \dv{\gamma^\mu}{s}\dd{s}\;,
\end{equation}
where $\D x^\mu \big|_\gamma$ is the embedding of the coordinate functions over the curve $\gamma$ and $\{a,b\}$ is the \emph{boundary} of $\gamma$, i.e. $\partial\gamma = \{a,b\}$. In general, a $p$-form is integrated over a $p$-chain, which is a map $[0,1]\times\stackrel{p}{\ldots}\times[0,1] \mapsto M$. The set of all $p$-chains is a vector space, which we denote as $\mathcal{C}_p$, and the boundary map is a linear map between these vector spaces, namely $\partial \vcentcolon \mathcal{C}_p \mapsto \mathcal{C}_{p-1}$. This map allows to define the \emph{homology} of $M$, i.e. the number of holes of the manifold. The $k$-th homology group is defined as the vector space of $k$-chains with no boundary (sometimes called $k$-cycles) which are not the boundary of a $k+1$-chain, namely as the quotient

\begin{equation}\label{homology group}
    H_k(M) \defeq \faktor{\left\{c\in \mathcal{C}_k\,\mid\, \partial c = 0\right\}}{\left\{c\in \mathcal{C}_k\,\mid\, c = \partial c^\prime \,,\, c^\prime \in \mathcal{C}_{k+1}\right\}}\;.
\end{equation}
The dimension of $H_k(M)$ gives the number of $k$-dimensional holes of $M$. The notion of boundary is somewhat the opposite of the notion of exterior derivative. Indeed, while the boundary decreases by one the degree of chains, the exterior derivative goes in the opposite direction, increasing by one the dimension of forms. The exterior derivative allows one to define the cohomology of a manifold, indeed the $k$-cohomology group is defined as the vector space of closed $p$-forms $(\D \omega = 0)$ which are not exact ($\omega \neq \D\alpha$), namely it is the quotient

\begin{equation}\label{cohomology group}
    H^k(M) \defeq \faktor{\left\{\omega\in \Omega^{k}(M,\mathbb{R})\,\mid\, \D\omega = 0\right\}}{\left\{\omega\in \Omega^k(M,\mathbb{R}),\mid\, \omega = \D \omega^\prime \,,\, \omega^\prime \in \Omega^{k-1}(M,\mathbb{R})\right\}}\;.
\end{equation}
The De Rham theorem states that \eqref{homology group} and \eqref{cohomology group} are isomorphic, $H_k(M)\cong H^k(M)$, therefore also the dimension of $H^k(M)$ gives the number of $k$-dimensional holes of $M$. The isomorphism is given by the integral which to a given $k$-chain, $c$, associates the unique $k$-form $\omega$ such that the integral of $\omega$ over $c$ is $0$

\begin{equation}
    c\mapsto \omega \;\;\;\Big| \;\;\; \int_c \omega = 0\;.
\end{equation}
The integral therefore defines a product between $k$-chains and $k$-forms, $\int_c \omega \defeq (c,\omega)$.
Finally, through the Stokes theorem, a duality between the boundary and the exterior derivative is established

\begin{equation}\label{Stoke}
    (\D \omega, c) = \int_c \D\omega = \int_{\partial c} \omega = (\omega, \partial c)\;,
\end{equation}
where $c\in \mathcal{C}_{k+1}$ and $\omega \in \Omega^{k}(M,\mathbb{R})$.

Having reviewed the relevant notions of spacetime calculus, we can now move to field-space calculus. For a thorough presentation of the topics just discussed, with several worked out examples, the reader is referred to \cite{nakahara2003geometry}.

\subsubsection{Field space (variational) calculus} We can define analogous operations on the space of fields. The field space $\Gamma$ is defined as the set of all possible fields configurations and it is assumed to be a differentiable manifold. If this is the case, it is possible to introduce a calculus on the space of forms on $\Gamma$. A $1$-form is an element of $T^*\Gamma$ which maps vector fields in $T\Gamma$ to functionals in $F=C^\infty(\Gamma)$. A $p$-form is an element of $\bigwedge\nolimits^{p} T^*\Gamma$ and the space of all forms is given by

\begin{equation}
    \Omega^\bullet(\Gamma,F)\defeq \bigoplus_{p=0}^{\text{dim}\,\Gamma}\bigwedge\nolimits^{p} T^*\Gamma\;,
\end{equation}
which is called \emph{variational complex}. $0$-forms are now functionals of the fields. The exterior derivative and the interior product are maps

\begin{eqnarray}
    &&\delta \vcentcolon \Omega^p(M,F) \mapsto \Omega^{p+1}(M,F)\;,\label{field space derivative}\\
    &&I_{\hat{V}} \vcentcolon \Omega^p(M,F) \mapsto \Omega^{p-1}(M,F)\;,
\end{eqnarray}
where $\hat{V}\in T\Gamma$. The exterior derivative increases by one the degree of a form and it is commonly referred to as field variation, while the interior product decreases by one the degree of a form and it is usually identified with a field contraction. The properties of these two maps are similar to the ones we have in spacetime calculus: the exterior derivative is a nilpotent operator, $\delta^2=0$, the interior product of a functional is $0$ and two contractions anti-commute. It is also assumed, unless otherwise stated, that vector fields in $TM$ are fields-independent, namely $\fL_{\hat{V}} X = 0\;\,\forall X\in TM,\hat{V}\in T\Gamma$, and $\fL_{\hat{V}}$ the field space Lie derivative defined hereafter. The exterior derivative and interior product define the field space Lie derivative through the analogue of Cartan's magic formula

\begin{equation}\label{Lie der field space}
    \fL_{\hat{V}} \defeq \delta I_{\hat{V}} + I_{\hat{V}} \delta\;.
\end{equation}
Of course, the Lie derivative does not change the degree of forms. The De Rham complex and the variational complex can be put together yielding the \emph{variational bi-complex} defined on $(M,\Gamma)$. The forms on the variational bi-complex will be denoted as $(p,q)$-forms. A $(p,q)$-form is a $p$-form in spacetime and a $q$-form in field space. The exterior derivative on the bi-complex is given by $\D+\delta$. There are mainly two conventions that can be adopted for this derivative:

\begin{itemize}
    \item We can assume that $\D+\delta$ is a coboundary, namely that $(\D+\delta)^2=0$, which yields $\D\delta=-\delta\D$;
    \item We can assume that $\D$ and $\delta$ commute, namely $\D\delta = \delta\D$.
\end{itemize}
The two conventions above yield the same results when equations are evaluated on shell of the equations of motion. While the first convention has been adopted in \cite{Ciambelli:2022vot}, we will adopt the second convention in these notes.

In field space it is possible to define coordinate functions as done in spacetime. If $\phi^i(x)$ are coordinates on $\Gamma$, then $\delta\phi^i(x)$ is a basis on $T^*\Gamma$ and $\fdv{\phi^i(x)}$, namely the functional derivatives with respect to coordinate functions on field space, is a basis of $T\Gamma$. A generic $1$-form on field space can therefore be written as

\begin{equation}
    \hat{\omega} = \int_M {\hat{\omega}}_i (x)\delta\phi^i(x)\dd^d x \defeq {\hat{\omega}}_i\delta\phi^i \; ,
\end{equation}
namely, the omitted sum in this case involves also an integration over spacetime (we can think of $x$ as another index on which we have to sum). For instance, a $(1,2)$-form is  written as

\begin{equation}
    \omega = \omega_{\mu,ij}\,\D x^\mu\otimes \delta\phi^i\wedge \delta\phi^j\;,
\end{equation}
and so on. In general, given a functional $f[\varphi] = \int_M h(\varphi(x),x)\dd[d]{x}$, with $h(\varphi(x),x)$ being a function depending on the spacetime $x$ and the field $\varphi$ evaluated at $x$, the functional derivative is given by

\begin{equation}
    \fdv{f[\varphi]}{\varphi} = \int_M \fdv{f[\varphi]}{\varphi(x)} \dd[d]{x}= \int_M\int_M \fdv{h(\varphi(y),y)}{\varphi(x)}\dd[d]{y}\dd[d]{x}=  \int_M \fdv{h(\varphi(x),x)}{\varphi(x)}\dd[d]{x}\;,
\end{equation}
since $\fdv{\phi(y)}{\phi(x)} = \delta^{(d)}(x-y)$.

\subsubsection{Lagrangian approach}
We want now to apply the above formalism to a Lagrangian theory. The action of a theory is given by

\begin{equation}
    S = \int_M L = \int_M \tilde{L}(\varphi,x)\sqrt{\abs{\det{g}}}\dd[d]{x}\;,
\end{equation}
where $L$ is a $(d,0)$-form and $\tilde{L}$ is a $(0,0)$-form (namely the functional of fields and their derivatives that multiplies the volume form). If we vary the Lagrangian with respect to the fields, it is always possible to rewrite such a variation as a term linear in the variation and a total derivative, namely 

\begin{equation}\label{eom and pre-symplectic lagrangian}
    \delta L = (\eom)\delta\varphi + \D\theta \he \D \theta\;,
\end{equation}
where $\eom$ stands for \emph{equations of motion}, $\he$ denotes an equality that is valid on shell of the equations of motion and $\theta$ is a $(d-1,1)$-form called \emph{pre-symplectic} potential. If $M$ has a boundary $\partial M$, this means that 

\begin{equation}\label{eom and pre-sympletic potential}
    \delta S = \int_M\Big[(\eom)\delta\varphi + \D\theta\Big] \he \int_{\partial M} \theta\;.
\end{equation}
From the pre-symplectic potential we can define the pre-symplectic local density

\begin{equation}
    \omega \defeq \delta \theta\;,
\end{equation}
which is a $(d-1,2)$-form, and from this, by integrating over a Cauchy slice $\Sigma$ (a codimension-1 submanifold on $M$), we define

\begin{equation}
    \Omega \defeq \int_\Sigma \omega = \int_\Sigma \delta\theta\;,
\end{equation}
which is a $(0,2)$-form called \emph{pre-symplectic $2$-form}. This is a key ingredient of this formalism since, as we will see, it encodes the Poisson brackets.

It is of course possible to define isometries on field space similarly to what has been done in spacetime. An isometry of $\omega$, for instance, is a vector field $\hat{V}$ such that 

\begin{equation}
    \fL_{\hat{V}} \omega = 0\;.
\end{equation}
Since $\omega = \delta \theta$ and $\delta^2 =0$, assuming a trivial cohomology on the space of $1$-forms over $\Gamma$ we have that 

\begin{equation}
    \fL_{\hat{V}} \omega = 0 \Rightarrow \delta I_{\hat{V}} \omega = 0 \Rightarrow I_{\hat{V}}\omega \defeq -\delta J_{\hat{V}}\;,
\end{equation}
where $J_{\hat{V}}$ is a $(d-1,0)$-form. Analogously, an isometry of $\Omega$ is defined as

\begin{equation}\label{symplectomorphism}
    \fL_{\hat{V}}\Omega = 0\;.
\end{equation}
A vector field satisfying \eqref{symplectomorphism} is called \emph{symplectomorphism}.
If we do not consider embedding fields in our theory, we can move the exterior derivative $\delta$ inside the integral, therefore $\delta \Omega = \delta\int_{\Sigma}\omega = \int_{\Sigma} \delta \omega = \int_\Sigma \delta^2\theta = 0$. Thus, given a symplectomorphism, we have

\begin{equation}
    0=\fL_{\hat{V}}\Omega = \delta I_{\hat{V}}\Omega\;,
\end{equation}
Since we have assumed trivial cohomology on the space of $1$-forms over $\Gamma$, the above equation yields 

\begin{equation}\label{hamiltonian vector field}
    I_{\hat{V}}\Omega \defeq -\delta H_{\hat{V}} \;\;,\;\; H_{\hat{V}} = \int_{\Sigma} J_{\hat{V}}\;,
\end{equation}
where $H_{\hat{V}}$ are called canonical charges and they are defined as the contraction of the pre-symplectic $2$-form. A vector field $\hat{V}\in T\Gamma$ satisfying \eqref{hamiltonian vector field} is called \emph{Hamiltonian vector field}. For trivial cohomology, Hamiltonian vector fields are symplectomorphisms and vice versa.

We are now ready to derive the Noether's theorems in this formalism. Indeed, in Noetherian jargon, $J_{\hat{V}}$ are called \emph{Noether currents}  while $H_{\hat{V}}$ are the \emph{Noether charges} and are the physical charges of the theory.

\subsection{Noether's theorems and charge algebra}\label{Noetherstheoremsandchargealgebra}

We have already introduced in the previous subsection the \textit{local} Noether current $J_{\hat{V}}$. Its expression is determined by the nature of the symmetry. In the case of an \textit{internal symmetry}, we have that
\begin{equation}
	\delta J_{\hat{V}} = - I_{\hat{V}}\omega \ ,
\end{equation}
and by taking into account that $\omega = \delta \theta$,
\begin{equation}
	\delta J_{\hat{V}} = - (\mathfrak{L}_{\hat{V}} \theta - \delta I_{\hat{V}} \theta) \ .
\end{equation}
In the following, we will assume that $\mathfrak{L}_{\hat{V}} \theta = 0$. While there are situations in which an internal symmetry acts non-covariantly on $\theta$, this assumption is fairly general, and allows us to simplify the discussion hereafter. We then  obtain
\begin{equation}
	J_{\hat{V}} = I_{\hat{V}} \theta \ ,
\end{equation}
modulo $\delta$-exact terms. This calculation is completely general for internal symmetries, where the equations of motion have not been imposed at all. Let us analyze the case of a \textit{spacetime symmetry}. In this case, we will consider that, associated to a vector field $\xi$ in spacetime, we have a vector field in field space denoted by $\hat{\xi}$. Then,
\begin{equation}
	\delta J_{\hat{\xi}} = - (\mathfrak{L}_{\hat{\xi}} \theta - \delta I_{\hat{\xi}} \theta) \ .
\end{equation}
As we are considering now spacetime transformations, $\mathfrak{L}_{\hat{\xi}} \theta = \mathcal{L}_{\xi} \theta$,\footnote{This is background independence, which is just $\delta_{\hat{\xi}} = \mathcal{L}_{\xi}$ in the standard formalism.} we get
\begin{equation}
	\delta J_{\hat{\xi}} = - (\iota_{\xi} \mathrm{d} \theta + \mathrm{d} \iota_{\xi} \theta - \delta I_{\hat{\xi}} \theta) \ .
\end{equation}	
If we assume that $\mathrm{d} \iota_{\xi} \theta$ vanishes at the boundary under consideration,\footnote{This assumption has to be reconsidered when there are fluxes, and led to different appreciations of the concept of Noether current in the literature.} and we go on-shell, we obtain
\begin{equation}
	J_{\hat{\xi}} \mathrel{\hat{=}} I_{\hat{\xi}} \theta - \iota_{\xi} L \ ,
\end{equation}
modulo $\delta$-exact terms. Unlike the case of internal symmetries, the equations of motion have been imposed. This current here derived is called the \textit{weakly-vanishing} Noether current, since it turns out to identically vanish on-shell, modulo $\mathrm{d}$-exact terms.

\subsubsection{Noether's first theorem}

This theorem concerns \textit{global symmetries}. It states that there is an associated codimension-1 conserved quantity for each global symmetry of the theory. This can be shown straightforwardly by evaluating the field space Lie derivative of the action, i.e
\begin{equation}
	\begin{split}
		\mathfrak{L}_{\hat{V}}S &= \int_M (I_{\hat{V}} \delta + \delta I_{\hat{V}}) L \\
		&= \int_M I_{\hat{V}} \delta L \\
		&= \int_M \left[ I_{\hat{V}} (\mathrm{e.o.m}) \delta \varphi + \mathrm{d} I_{\hat{V}} \theta \right] \ .
	\end{split}
\end{equation}
If $\hat{V}$ defines a global symmetry of the theory, then $\mathfrak{L}_{\hat{V}}S = 0$, and we further assume here $\mathfrak{L}_{\hat{V}}L = 0$. If we recall that $J_{\hat{V}}=I_{\hat{V}} \theta$, we finally get
\begin{equation}
	\mathrm{d} J_{\hat{V}} = - I_{\hat{V}} (\mathrm{e.o.m}) \delta \varphi \ ,
\end{equation}
hence
\begin{equation}
	\mathrm{d} J_{\hat{V}} \mathrel{\hat{=}} 0 \ .
\end{equation}
The result is that the local Noether current is conserved on-shell. As a consequence, the global functional defined as
\begin{equation}
	H_{\hat{V}} := \int_{\Sigma} J_{\hat{V}}
\end{equation}
is the integral of a codimension-1 conserved global charge on-shell. This is the associated Noether charge.

\subsubsection{Noether's second theorem}

This theorem concerns \textit{gauge symmetries}. It states that there is an associated codimension-2 conserved quantity for each gauge symmetry of the theory. Explicitly, the local Noether current is such that
\begin{equation}\label{noether second theorem}
	J_{\hat{V}} \mathrel{\hat{=}} \mathrm{d} Q_{\hat{V}} \ ,
\end{equation}
so it is a $\mathrm{d}$-exact term on-shell.\footnote{We are here considering $\hat{V}$ as a generic field space vector field. However, if $\hat{V}$ is associated to a spacetime diffeomorphism $\xi$, then the notation $\hat{\xi}$ is used in these notes.} Its conservation is directly inferred from $\mathrm{d}^2 = 0$. This can also be shown by evaluating the field space Lie derivative of the action, i.e
\begin{equation}
	\begin{split}
		\mathfrak{L}_{\hat{V}}S &= \int_M (I_{\hat{V}} \delta + \delta I_{\hat{V}}) L \\
		&= \int_M \left[ I_{\hat{V}} (\mathrm{e.o.m}) \delta \varphi + \mathrm{d} I_{\hat{V}} \theta \right] \\
		&= \int_M I_{\hat{V}} (\mathrm{e.o.m}) \delta \varphi + \int_{\partial M} I_{\hat{V}} \theta \ .
	\end{split}
\end{equation}
As before, $\mathfrak{L}_{\hat{V}}S = 0$ if $\hat{V}$ defines a gauge symmetry of the theory. Additionally, if $\hat{V} = \hat{\xi}$ defines a spacetime symmetry, then we assume $\mathfrak{L}_{\hat{\xi}} \theta = \mathcal{L}_{\xi} \theta$, which, as per the internal-symmetry discussion above, confines our analysis to gauge symmetries acting covariantly on $\theta$. Note that
\begin{equation}
	\begin{split}
		\mathcal{L}_{\xi} S &= \int_M \mathcal{L}_{\xi} L \\
		&= \int_M \mathrm{d} \iota_{\xi} L \\
		&= \int_{\partial M} \iota_{\xi} L \ .
	\end{split}
\end{equation}
Finally,
\begin{equation}
 \int_{\partial M} I_{\hat{\xi}} \theta \mathrel{\hat{=}} \int_{\partial M} \iota_{\xi} L \ ,
\end{equation}
and we obtain
\begin{equation}
	J_{\hat{\xi}} \mathrel{\hat{=}} \mathrm{d} Q_{\hat{\xi}} \ .
\end{equation}
Analogously, the global functional defined as
\begin{equation}
	H_{\hat{\xi}} := \int_{\Sigma} J_{\hat{\xi}}
\end{equation}
is a conserved charge on-shell. This is the associated Noether charge when considering gauge symmetries. Note that it is now a codimension-2 quantity, if the hypersurface $\Sigma$ is such that its boundary $\partial \Sigma = S$ is a codimension-2 surface. If so,
\begin{equation}
	H_{\hat{\xi}} \mathrel{\hat{=}} \int_S Q_{\hat{\xi}} \ .
\end{equation}
This is the reason why in the literature the name \textit{surface charge} is typically used for the Noether charge for gauge symmetries. This charge is also called \textit{corner charge} since the surface $S$ is generically called \textit{corner}.

\subsubsection{Charge algebra}
Let us explain now why the adjective ``pre''-symplectic is used for the 2-form $\Omega$. Recall that
\begin{equation}
	\Omega = \int_{\Sigma} \omega = \int_{\Sigma} \delta \theta \ ,
\end{equation}
where $\omega$ and $ \theta$ are the pre-symplectic local density and potential respectively. A 2-form $\Omega'$ is called symplectic if it is characterized by the following properties:
\begin{itemize}
	\item $\delta \Omega' = 0$
	\item $ I_{\hat{V}} \Omega' = 0 \Longleftrightarrow \hat{V} = 0$
\end{itemize}
That is, the 2-form $\Omega'$ is closed in field space and non-degenerate. In this notes, the 2-form $\Omega$ is called pre-symplectic because the non-degeneracy property does not hold in general. This means that there are non-vanishing field space vector fields $\hat{V}$ (non-vanishing symmetries) such that $I_{\hat{V}} \Omega = 0$. If we quotient these symmetries out the pre-symplectic 2-form $\Omega$ becomes symplectic and can define a \textit{Poisson bracket} of the \textit{charge algebra}. This Poisson bracket is defined as
\begin{equation}
	\lbrace H_{\hat{V}} , H_{\hat{W}} \rbrace = \mathfrak{L}_{\hat{W}} H_{\hat{V}} \ ,
\end{equation}
where $\hat{V} , \hat{W} \in T \Gamma$ are symplectomorphisms. Indeed, one is able to show that the Poisson bracket is skew-symmetric by using the properties of symplectomorphisms. Explicitly,
\begin{equation}
	\begin{split}
		\lbrace H_{\hat{V}} , H_{\hat{W}} \rbrace &= I_{\hat{W}} \delta H_{\hat{V}} \\
		&= - I_{\hat{W}} I_{\hat{V}} \Omega \\
		&= I_{\hat{V}} I_{\hat{W}} \Omega \\
		&= - I_{\hat{V}} \delta H_{\hat{W}} \\
		&= - \lbrace H_{\hat{W}} , H_{\hat{V}} \rbrace \ .
	\end{split}
\end{equation}
Apart from proving its skew-symmetry, we also showed that the Poisson bracket is determined by the symplectic 2-form $\Omega$, i.e
\begin{equation}
	\lbrace H_{\hat{V}} , H_{\hat{W}} \rbrace = I_{\hat{V}} I_{\hat{W}} \Omega \ .
\end{equation}

At this point, we consider spacetime vector fields $\xi , \zeta \in T M$ associated with spacetime symmetries. The corresponding Poisson bracket reads
\begin{equation}
	\lbrace H_{\hat{\xi}} , H_{\hat{\zeta}} \rbrace = \mathfrak{L}_{\hat{\zeta}} H_{\hat{\xi}} \ .
\end{equation}
In this case, we also have the Lie bracket vector fields in spacetime, which reads
\begin{equation}
	[\xi , \zeta] = \mathcal{L}_\xi \zeta \ .
\end{equation}
One may now wonder how this Lie bracket of (spacetime) vector fields defining the symmetry algebra is related to the Poisson bracket of the charges associated to the corresponding (field space) vector fields defining the charge algebra. This is a fundamental question that we will answer next. Let us start by writing the identity
\begin{equation}
	I_{[\hat{\xi} , \hat{\zeta}]} = [\mathfrak{L}_{\hat{\xi}} , I_{\hat{\zeta}}] \ .
\end{equation}
If we apply it to the symplectic 2-form $\Omega$, we get
\begin{equation}
	I_{[\hat{\xi} , \hat{\zeta}]} \Omega = \mathfrak{L}_{\hat{\xi}} I_{\hat{\zeta}} \Omega - I_{\hat{\zeta}} \mathfrak{L}_{\hat{\xi}} \Omega \ .
\end{equation}
Since $\hat{\xi}$ is a symplectomorphism, $\mathfrak{L}_{\hat{\xi}} \Omega = 0$, and we have
\begin{equation}
	I_{[\hat{\xi} , \hat{\zeta}]} \Omega = \delta I_{\hat{\xi}} I_{\hat{\zeta}} \Omega + I_{\hat{\xi}} \delta I_{\hat{\zeta}} \Omega \ .
\end{equation}
If we recall that $\delta \Omega = 0$, the second term of the above equation gives $I_{\hat{\xi}} \delta I_{\hat{\zeta}} \Omega = I_{\hat{\xi}} \mathfrak{L}_{\hat{\zeta}} \Omega = 0$, since also $\hat{\zeta}$ is a symplectomorphism. Therefore,
\begin{equation}
	I_{[\hat{\xi} , \hat{\zeta}]} \Omega = \delta I_{\hat{\xi}} I_{\hat{\zeta}} \Omega \ ,
\end{equation}
and thus, introducing the Poisson bracket,
\begin{equation}
	I_{[\hat{\xi} , \hat{\zeta}]} \Omega = \delta \lbrace H_{\hat{\xi}} , H_{\hat{\zeta}} \rbrace \ .
	\label{Eq1}
\end{equation}
Although we will not demonstrate it explicitly in these notes, one can see using similar arguments that the Lie brackets Jacobi identity induces the Poisson brackets Jacobi identity.

Finally, we can derive the correspondence between the charge algebra given by the Poisson brackets and the symmetry algebra given by the Lie brackets. First, note that the identity
\begin{equation}
	I_{\hat{\xi}} \Omega = - \delta H_{\hat{\xi}}
\end{equation}
relates the Noether charge $H_{\hat{\xi}}$ and the field space vector field $\hat{\xi}$ associated to the spacetime symmetry $\xi$ via the symplectic 2-form $\Omega$. It we consider two charges, for instance $H_{\hat{\xi}}$ and $H_{\hat{\xi}} + \kappa$, both charges correspond to the same vector field $\hat{\xi}$ as long as $\delta \kappa = 0$. This correspondence is said to be cohomological. We can similarly conclude from Eq.~\eqref{Eq1} that both $\lbrace H_{\hat{\xi}} , H_{\hat{\zeta}} \rbrace$ and $\lbrace H_{\hat{\xi}} , H_{\hat{\zeta}} \rbrace + \kappa_{\hat{\xi} , \hat{\zeta}}$ correspond to the same vector field $[\hat{\xi} , \hat{\zeta}]$ as long as $\delta \kappa_{\hat{\xi} , \hat{\zeta}} = 0$. Finally, we can write
\begin{equation}
	I_{[\hat{\xi} , \hat{\zeta}]} \Omega = - \delta H_{[\hat{\xi} , \hat{\zeta}]}\ ,
\end{equation}
to deduce from \eqref{Eq1} and the above cohomological argument that
\begin{equation}\label{projectiverepresentation}
	\lbrace H_{\hat{\xi}} , H_{\hat{\zeta}} \rbrace = - H_{[\hat{\xi} , \hat{\zeta}]} + \kappa_{\hat{\xi} , \hat{\zeta}} \ ,
\end{equation}
with $\delta \kappa_{\hat{\xi} , \hat{\zeta}} = 0$. We can read from this result that the Poisson bracket of charges represents the Lie bracket of symmetries \textit{projectively}. This is essentially the same as stating that the charge algebra represents the symmetry algebra  modulo \textit{central extensions}. A central extension is just a constant in field space appearing in the bracket structure of a theory. It can be considered as a new algebra generator, with the word ``central'' characterizing its commutativity with the rest of the algebra generators. This is a general and highly important result, in particular for the theory of asymptotic symmetries that we will introduce. We will do it in \cref{Sec1.4}, but in \cref{Sec1.3} we will first apply the formalism presented in previous subsections to the computation of conserved charges in gauge theories.

\subsection{Application to gauge theories} \label{Sec1.3}

Let us consider a classical gauge theory on a 4-dimensional flat spacetime. We are generally interested in non-abelian gauge theories, with compact simple gauge Lie group $G$ and corresponding Lie algebra $\mathfrak{g}$. The Lie algebra generators $T_a$ satisfy
\begin{equation}
	\left[T_a, T_b\right] = f_{ab}^c T_c \ ,
\end{equation}
where $f_{ab}^c$ are the structure constants.\footnote{Note that the structure constants are antisymmetric in the indices $a$, $b$.} If the group $G$ is a compact semi-simple Lie group, it is possible to choose the Lie algebra generators so that they satisfy the orthogonality condition\footnote{The structure constants obtained as $f_{abc} = f_{ab}^d \delta_{dc}$ are completely antisymmetric in the indices $a$, $b$, $c$.}
\begin{equation}
	\mathrm{Tr} ( T_a T_b ) = \frac{\delta_{ab}}{2} \ .
\end{equation}
The gauge connection/field $A$ is a Lie algebra-valued 1-form $A = A_{\mu}^a \, \mathrm{d} x^{\mu} \otimes T_a$, and the field strength $F$ is a Lie algebra-valued 2-form $F = \frac{1}{2} F_{\mu \nu}^a \mathrm{d}x^{\mu} \wedge \mathrm{d}x^{\nu} \otimes T_a$ given by
\begin{equation}
	F = \mathrm{d} A + A \wedge A \ .
\end{equation}
Recall that $F_{\mu \nu}^a = \delta_{\mu \nu}^{\rho \sigma} \partial_{\rho} A_{\sigma}^a + f_{bc}^a A_{\mu}^b A_{\nu}^c$, with $\delta$ the generalized Kronecker tensor. We also introduce the gauge-covariant exterior derivative,
\begin{equation}
	\mathrm{D} \equiv \mathrm{d} + A \ ,
\end{equation}
as the generalization of the exterior derivative $\mathrm{d}$ (c.f. \eqref{exterior derivative}) with respect to the gauge connection $A$.

The Lagrangian reads in this setting
\begin{equation}
	L = \frac{1}{2} \mathrm{Tr} \left( F^2 \right) \, \mathrm{Vol}_M  \ ,
\end{equation}
where $\mathrm{Vol}_M$ is the volume form of the manifold $M$. This volume form $\mathrm{Vol}_M$ is, by definition, a top-form, then $\mathrm{Vol}_M = \star 1$, with $\star$ denoting the Hodge duality operator. If we consider its definition in \eqref{star hodge}, we have
\begin{equation}
	\mathrm{Vol}_M = \frac{1}{4!} \epsilon_{\alpha \beta \gamma \delta} \, \mathrm{d}x^{\alpha} \wedge \mathrm{d}x^{\beta} \wedge \mathrm{d}x^{\gamma} \wedge \mathrm{d}x^{\delta} \ ,
\end{equation}
where $\epsilon_{\alpha \beta \gamma \delta}$ is the covariant Levi-Civita tensor in flat spacetime (which is equal to the Levi-Civita symbol). Its contravariant version $\epsilon^{\alpha \beta \gamma \delta}$ is defined by means of the metric as usual, so we can now use the identity
\begin{equation}
	\mathrm{d}x^{\alpha} \wedge \mathrm{d}x^{\beta} \wedge \mathrm{d}x^{\gamma} \wedge \mathrm{d}x^{\delta} = - \epsilon^{\alpha \beta \gamma \delta} \, \mathrm{d}x^0 \wedge \mathrm{d}x^1 \wedge \mathrm{d}x^2 \wedge \mathrm{d}x^3
\end{equation}
to write
\begin{equation}
	\mathrm{Vol}_M = \frac{1}{4!} \delta_{\alpha \beta \gamma \delta}^{\alpha \beta \gamma \delta} \, \mathrm{d}x^0 \wedge \mathrm{d}x^1 \wedge \mathrm{d}x^2 \wedge \mathrm{d}x^3 \ ,
\end{equation}
but $\delta_{\alpha \beta \gamma \delta}^{\alpha \beta \gamma \delta} = 4!$, so
\begin{equation}
	\mathrm{Vol}_M = \mathrm{d}x^0 \wedge \mathrm{d}x^1 \wedge \mathrm{d}x^2 \wedge \mathrm{d}x^3 \ .
\end{equation}
The above relation and the skew-symmetry of the strength tensor allow us to write the Lagrangian as
\begin{equation}
	\begin{split}
			L &= \frac{1}{2} \mathrm{Tr} \left( F^2 \right) \, \mathrm{Vol}_M \\
			&= - \frac{1}{4} F_{\mu \nu}^a F_a^{\mu \nu} \, \mathrm{d}x^0 \wedge \mathrm{d}x^1 \wedge \mathrm{d}x^2 \wedge \mathrm{d}x^3 \\
			&= - \frac{1}{8} F_{\mu \nu}^a \delta_{\rho \sigma}^{\mu \nu} F_a^{\rho \sigma} \, \mathrm{d}x^0 \wedge \mathrm{d}x^1 \wedge \mathrm{d}x^2 \wedge \mathrm{d}x^3 \\
			&= \frac{1}{16} F_{\mu \nu}^a \epsilon_{\alpha \beta \rho \sigma} \epsilon^{\alpha \beta \mu \nu} F_a^{\rho \sigma} \, \mathrm{d}x^0 \wedge \mathrm{d}x^1 \wedge \mathrm{d}x^2 \wedge \mathrm{d}x^3 \\
			&= - \frac{1}{16} F_{\mu \nu}^a \epsilon_{\alpha \beta \rho \sigma} F_a^{\rho \sigma} \, \mathrm{d}x^{\alpha} \wedge \mathrm{d}x^{\beta} \wedge \mathrm{d}x^{\mu} \wedge \mathrm{d}x^{\nu} \\
			&= - \frac{1}{4} \left( \frac{1}{2} F_{\mu \nu}^a \mathrm{d}x^{\mu} \, \wedge \mathrm{d}x^{\nu} \right) \wedge \left( \frac{1}{2} \epsilon_{\alpha \beta \rho \sigma} F_a^{\rho \sigma} \mathrm{d}x^{\alpha} \wedge \mathrm{d}x^{\beta} \right) \\
			&= - \frac{1}{2} \mathrm{Tr} ( F \wedge \star F ) \ .
		\end{split}
\end{equation}

The gauge connection is the dynamical field when considering gauge theories in flat spacetime. The field space $\Gamma$ is then formed by $A$. Its equations of motion and the pre-symplectic potential follow from taking the corresponding field variation {(c.f. \eqref{field space derivative}) of the action,
\begin{equation}
	\delta S = \int_M \delta L \ ,
\end{equation}
giving
\begin{equation}
	\begin{split}
			\delta S &= - \frac{1}{2} \int_M \mathrm{Tr} \, \delta ( F \wedge \star F ) \\
			&= - \int_M \mathrm{Tr} ( \delta F \wedge \star F ) \\
			&= - \int_M \mathrm{Tr} \left( ( \mathrm{d} \delta A + A \wedge \delta A + \delta A \wedge A ) \wedge \star F \right) \\
			&= \int_M \mathrm{Tr} \left( ( \mathrm{d} \star F + [A, \star F] ) \wedge \delta A - \mathrm{d} ( \delta A \wedge \star F ) \right) \\
			&= \int_M \mathrm{Tr} \left( \mathrm{D} \star F \wedge \delta A - \mathrm{d} ( \delta A \wedge \star F ) \right) \ .
	\end{split}
\end{equation}
where we have used $\delta \mathrm{d} = \mathrm{d} \delta$. We can now read off the field equations for $A$,
\begin{equation}
	\mathrm{D} \star F = 0 \ ,
\end{equation}
which are the Yang-Mills equations (together with the Bianchi identity $\mathrm{D} F = 0$) as expected, as well as the pre-symplectic potential (see \eqref{eom and pre-symplectic lagrangian} and \eqref{eom and pre-sympletic potential}),
\begin{equation}
	\theta = - \, \mathrm{Tr} ( \delta A \wedge \star F ) \ .
\end{equation}

The corresponding gauge symmetry of the theory is encoded in the invariance under the well-known transformation for a gauge field $I_{\hat{\lambda}} \delta A :=\delta_{\lambda} A   = \mathrm{D} \lambda = \dd \lambda + [A, \lambda]$, with $[A, \lambda] = f_{a b}^c A^a \lambda^b T_c$ the Lie algebra commutator. Note that
\begin{equation}
	\begin{split}
			\fL_{\hat\lambda} F :&= I_{\hat{\lambda}} \delta F \\
			&= I_{\hat{\lambda}} ( \mathrm{d} \delta A + A \wedge \delta A + \delta A \wedge A ) \\
			&= I_{\hat{\lambda}} \mathrm{D} \delta A \\
			&= \mathrm{D}^2 \lambda \\
			&= ( \mathrm{d} + A ) ( \mathrm{d} \lambda + [A, \lambda] ) \\
			&= \mathrm{d} [A, \lambda] + [A, \mathrm{d} \lambda] + [A , [A, \lambda]] \\
			&= [\mathrm{d} A + A \wedge A, \lambda] \\
			&= [F, \lambda] \ ,
		\end{split}
\end{equation}
hence
\begin{equation}
	\begin{split}
		\fL_{\hat\lambda} S = - \int_M \mathrm{Tr} ( [F, \lambda] \wedge \star F )= 0 \;,
	\end{split}
\end{equation}
because, in components,
\begin{equation}
	\begin{split}
		\mathrm{Tr} ( [F, \lambda] \wedge \star F ) \; \propto\; \delta_{a b} f_{c d}^a F_{\mu \nu}^c \lambda^d F^{\mu \nu b} = 0 \ ,
	\end{split}
\end{equation}
since $\delta_{a b} f_{c d}^a = f_{b c d}$ is antisymmetric in the indices $b, c$, while $F_{\mu \nu}^c F^{\mu \nu b}$ is symmetric. Therefore, our theory is indeed invariant under transformations $\delta_{\lambda} A   = \mathrm{D} \lambda$. 

The associated Noether current follows from evaluating (see \eqref{hamiltonian vector field})
\begin{equation}
	I_{\hat{\lambda}} \Omega = \int_{\Sigma} I_{\hat{\lambda}} \omega = \int_{\Sigma} I_{\hat{\lambda}} \delta \theta \ ,
\end{equation}
giving
\begin{equation}
	\begin{split}
			I_{\hat{\lambda}} \Omega &= - \int_{\Sigma} \mathrm{Tr} [ I_{\hat{\lambda}} ( \delta A \wedge \star \delta F ) ] \\
			&= - \int_{\Sigma} \mathrm{Tr} ( \mathrm{D} \lambda \wedge \star \delta F - \delta A \wedge [\star F, \lambda] ) \\
			&= - \int_{\Sigma} \mathrm{Tr} ( \mathrm{d} \lambda \wedge \star \delta F + [A, \lambda] \wedge \star \delta F - \delta A \wedge [\star F, \lambda] ) \\
			&= - \int_{\Sigma} \mathrm{Tr} \, \delta ( \mathrm{d} \lambda \wedge \star F + [A, \lambda] \wedge \star F) \; ,
		\end{split}
\end{equation}
where we used $\delta\lambda=0$. The Noether current is then,
\begin{equation}
	J_{\hat{\lambda}} = - \, \mathrm{Tr} ( \mathrm{d} \lambda \wedge \star F + [A, \lambda] \wedge \star F ) \ .
\end{equation}
It is possible to rewrite this Noether current as an on-shell total derivative, due to Noether's second theorem (c.f. \eqref{noether second theorem}). Explicitly,
\begin{equation}
	\begin{split}
			J_{\hat{\lambda}} &= - \, \mathrm{Tr} ( \mathrm{d} \lambda \wedge \star F + [A, \lambda] \wedge \star F  ) \\
			&= - \, \mathrm{Tr} \left( \mathrm{d} ( \lambda \star F ) - \lambda ( \mathrm{d} \star F + [A, \star F] ) \right) \\
			&= - \, \mathrm{Tr} \left( \mathrm{d} ( \lambda \star F ) - \lambda \mathrm{D} \star F \right) \\
			&\mathrel{\hat{=}} - \, \mathrm{Tr} \left( \mathrm{d} ( \lambda \star F ) \right) \ .
		\end{split}
\end{equation}
Finally, the Noether charge (c.f. \eqref{hamiltonian vector field}) is given by
\begin{equation}
	H_{\hat{\lambda}} = \int_{\Sigma} J_{\hat{\lambda}} \ ,
\end{equation}
hence
\begin{equation}
	\begin{split}
			H_{\hat{\lambda}} &\mathrel{\hat{=}} - \int_{\Sigma} \mathrm{Tr} \left( \mathrm{d} (\lambda \star F) \right) \\
			&= - \int_S \mathrm{Tr} ( \lambda \star F ) \ ,
		\end{split}
\end{equation}
where $\Sigma$ is an arbitrary Cauchy surface with boundary $S$. 

This is our final result,  showing how Noether's second theorem applies to gauge theories and how to explicitly compute the  conserved charge associated to the local  symmetry of the theory. The charge has support on a codimension-2 surface $S$, called the corner. We now turn our attention to the aforementioned theory of asymptotic symmetries.

\subsection{Asymptotic symmetries}\label{Sec1.4}

What we have learned so far is that the basic ingredients needed to define a classical physical theory are the spacetime manifold $M$ with boundary $B$ and a field space $\Gamma$. These are the basic ingredients of the theory of asymptotic symmetries. 

A classical dynamical theory is specified by
\begin{itemize}
    \item[(i)]  The \textbf{dynamics} on $M$, i.e. the Lagrangian describing the system.
    \item[(ii)] \textbf{Boundary conditions} $\Gamma\rvert_B$ defining the asymptotic behaviour of the bulk fields near the boundary, namely defining the so-called \emph{falloffs} of the bulk fields.
    \item[(iii)] \textbf{Gauge fixing} conditions if the theory is a gauge theory.
\end{itemize}
At this point, in order to compute the Noether charges we need to identify the real physical symmetries of the system. To this aim, let us introduce the definitions of residual symmetries and trivial symmetries.
\begin{itemize}
    \item \textbf{Residual symmetries}: they are defined as symplectomorphisms preserving the dynamics in the bulk and the boundary conditions. We stress that they are usually known as \emph{gauge transformations}.
    \item \textbf{Trivial symmetries}: they are residual symmetries with vanishing Noether charges and therefore they are true redundancies of the system. Strictly speaking, these are the quantities that should be called gauge symmetries.
    \item \textbf{Asymptotic symmetries}: they are residual symmetries with non-vanishing Noether charges. They are, therefore, physical transformations acting non-trivially on the field space, mapping the system into an inequivalent configuration.
\end{itemize}
Due to their redundancy nature, the trivial transformations represent the zero modes of the pre-symplectic 2-form. Additionally, the group they form is an ideal inside the residual symmetry group. Finally, the only symmetries to focus on are the non-trivial residual symmetries that we have defined as asymptotic symmetries. Precisely, the asymptotic symmetry group is defined as the quotient
\begin{equation}
    \text{Asymptotic symmetries}=\faktor{\text{Residual symmetries}}{\text{Trivial symmetries}}\;.
\end{equation}
Having restricted to the asymptotic symmetries, the pre-symplectic 2-form is now invertible and this allows us to deal with a well-defined Poisson bracket of physical charges. This defines the charge algebra of the theory and the physical observables consequently.

Note that to be as general as possible, in presence of a gauge theory the Noether charges \emph{should be} computed without any a priori restrictions given by gauge fixing. In fact, only after the charges' computation, the transformation needed to perform the gauge fixing \emph{should be} shown to be a trivial symmetry (leading to a vanishing charge). Typically, however, it is not possible to carry on this procedure due to technical reasons. Nevertheless, on the contrary, the gauge fixing procedure, as well as the enforcing of the boundary conditions, may lead to pathologies for the well-definiteness of physical charges. For this reason, the Noether charges associated to asymptotic symmetries are canonical if they satisfy these three conditions:
\begin{itemize}
    \item \textbf{Integrable}: $I_{\hat{V}} \Omega=-\delta \int_S Q_{\hat{V}}$. 
    \item \textbf{Conserved}: $ H_{\hat{V}}\rvert_{S_2}-H_{\hat{V}}\rvert_{S_1}=\int^{S_2}_{S_1} \dd Q_{\hat{V}}\he 0$.
    \item \textbf{Finite}: charges should not diverge approaching the codimension-2 surface $S$.
\end{itemize}
If these conditions are satisfied, points (ii) and (iii) are pathologies-free and the theory is well defined.

The last thing to mention before going on is that the asymptotic symmetry group can be (and typically is) larger than the bulk symmetry group. A remarkable example is given by Minkowski spacetime in $4$ dimensions, whose bulk symmetry group is given by the Poincaré group, while the asymptotic symmetry group at null infinity is given by the infinite dimansional Bondi-Van der Burg–Metzner–Sachs ($\mathrm{BMS_4}$) group that we will discuss in the next section.

\section{Gravity}\label{gravity sec}

\subsection{What is wrong?}\label{Sec2.1}

For the sake of simplicity, the previous discussion was done assuming a closed system. Then, as expected, we obtained conserved charges. We elaborate here on the problems one encounters when dealing with gravity. In this case, every one of the three previously-mentioned properties (integrability, conservation and finiteness) fails to be a priori true. 

\paragraph{Integrability} It is not always guaranteed that $I_{\hat{\xi}} \Omega=-\delta H_{\hat{\xi}}$. If this does not hold, then the charge algebra does not close and thus one cannot derive the Poisson bracket of the theory. Generally, we have 
\begin{align}
    I_{\hat{\xi}} \Omega=\int_{\Sigma}  I_{\hat{\xi}} \delta \theta=\int_{\Sigma} \mathrm{d}(\iota_{\xi}\theta+Q_{\delta \hat{\xi}})-\delta \mathrm{d}Q_{\hat{\xi}}\;,
\end{align}
where we added the term $Q_{\delta \hat{\xi}}$ in case $\xi$ is field dependant, but we will assume it is zero from now on. The term $\mathrm{d}(\iota_{\hat{\xi}}\theta)$ is not generically of the form $\delta(...)$, and therefore the total expression cannot be equal to $-\delta H_{\hat{\xi}}$. We can then define the \textit{symplectic flux} as
\begin{equation}
    F_{\hat{\xi}}=\int_S \iota_{\xi}\theta\;,
\end{equation}
where $S=\partial \Sigma$. This term is responsible for dissipation in the system.

There are various proposals to deal with non-integrability, and we will mention two of them:
\begin{enumerate}
    \item Following \cite{Barnich:2011mi} (see also \cite{Freidel:2021fxf}) we can split between integrable and non-integrable parts introducing a modified Poisson bracket. There is no canonical way to perform such splitting, and here we will use the Noetherian split proposed in \cite{Freidel:2021fxf}. This means that we split 
    \begin{equation}
        I_{\hat{\xi}} \Omega=-\delta H_{\hat{\xi}}+F_{\hat{\xi}}\;,
    \end{equation}
such that $H_{\hat{\xi}}=\int_S Q_{\hat{\xi}}$, with $\mathrm{d}Q_{\hat{\xi}}= I_{\hat{\xi}} \theta - \iota_{\xi} L$ being the unmodified Noether charge, and $F_{\hat{\xi}}=\int_S\iota_{\xi}\theta$ the Noetherian flux. It can then be proved that we have the modified bracket
\begin{equation}\label{modiBR}
     \poissonbracket{H_{\hat{\xi}}}{H_{\hat{\zeta}}}=\fL_{\hat{\zeta}} H_{\hat{\xi}}-I_{\hat{\zeta}} F_{\hat{\xi}} +\int_S i_\xi i_\zeta L\;.
\end{equation}
    \item We can consider an enlarged field space, and define another symplectic form $\Omega^{\mathrm{ext}}$ such that 
    \begin{equation}
        I_{\hat{\xi}} \Omega^{\mathrm{ext}}=-\delta H_{\hat{\xi}}\;, 
    \end{equation}
    for every diffeomorphism $\xi$. This procedure goes under the name of \emph{extended phase space} \cite{Ciambelli:2021vnn, Ciambelli:2021nmv, Freidel:2021dxw, Speranza:2022lxr}, and we will have to say more about it later. The  system is still dissipative but is now integrable, thanks to the introduction of edge modes\footnote{With ``edge mode'' we denote a field living on a codimension-2 surface.}. 
\end{enumerate}

\paragraph{Conservation} Charges are not always conserved. In general
\begin{equation}
    H_{\hat{\xi}}|_{S_2}- H_{\hat{\xi}}|_{S_1}=\int_{S_1}^{S_2}\mathrm{d}Q_{\hat{\xi}}=\int_{S_1}^{S_2}  I_{\hat{\xi}} \theta - \iota_{\xi} L
\end{equation}
is not equal to zero. There are two main reasons why this might happen. One is that there might be gravitational fluxes through the boundary under consideration. In other words, the  subregion we consider is not isolated from its complement (there is a ``leakage''). Another reason is that, for a gravitational theory defined on an odd-dimensional bulk with cosmological constant, a conformal anomaly exists, and this anomaly contributes to the non-conservation of charges (see \cite{Alessio:2020ioh}). This breaking of conformal symmetry is holographically understood as an anomalous Ward-Takahashy identity, as discussed below.  

\paragraph{Finiteness} The charges are guaranteed to be finite only if the boundary is at finite distance in the bulk. In gravity, however, we often deal with asymptotic boundaries.
Consider the case of gravity with a negative cosmological constant in three dimensions, with action (in suitable units) given by 
\begin{equation}
    S=\frac{1}{16\pi G}\int \mathrm{d}^3 x\;\sqrt{|\det{g} | }(R-2\Lambda)\;,
\end{equation}
where $\Lambda$ is a cosmological constant. Einstein equations imply $R=6\Lambda$, and the on-shell action is thus proportional to the spacetime volume, and therefore suffers from divergences.\footnote{We do not consider the Gibbons-Hawing-York term in this lecture notes, but it also plays an important role in the value of the on-shell action.} It is then clear that the surface charges, derived from this action, may also be infinite. We therefore have to perform a suitable renormalization, called \textit{phase space renormalization}. We add a boundary term to the Lagrangian 
\begin{equation}
    L_R=L+\mathrm{d}\ell_{\mathrm{div}}
\end{equation}
and tune it so to cure the divergences. Also, we add corner symplectic potentials $\nu_{\mathrm{div}}$, such that 
\begin{equation}
    \theta_R=\theta-\mathrm{d} \nu_{\mathrm{div}}-\delta\ell_{\mathrm{div}}\;.
    \end{equation}
This implies that we have 
\begin{equation}
    H_{\hat{\xi}}^R=H_{\hat{\xi}}-\int_S J_{\hat{\xi}}^{\mathrm{div}}\;,
\end{equation}
where $J_\xi^{\mathrm{div}}=I_{\xi}\nu_{\mathrm{div}}-\iota_\xi\ell_{\mathrm{div}}$. It is not a priori obvious that $\ell_\mathrm{div}$ and $\nu_{\mathrm{div}}$ can always be selected to cure divergences, but  this turns out to be true on a case by case basis. The symplectic flux also gets renormalised, and the final result is 
\begin{equation}
    F_{\hat{\xi}}^R=F_{\hat{\xi}}+\int_S (\delta \iota_\xi \ell_{\mathrm{div}}-\fL_{\hat\xi}\;\nu_{\mathrm{div}})\;.
\end{equation}
This procedure is the rigorous generalization of the addition of counterterms in AdS holographic renormalization.

Before providing a detailed gravitational example, we continue discussing $3$-dimensional gravity further. We saw earlier that Poisson brackets are a projective representation of Lie brackets. Brown and Henneaux showed in \cite{Brown:1986nw} that the asymptotic symmetry algebra of $AdS_3$ is given by two copies of the Virasoro algebra with equal central charges $c=\frac{3l}{2G}$, where $l$ is the $AdS_3$ radius, and $G$ the Newton constant. Nowadays, we understand this result in the light of holography ($AdS/CFT$ correspondence), \cite{Maldacena:1997re}. This correspondence establishes a duality between a gravitational theory defined on an (asymptotically) $AdS$ spacetime in the bulk and a conformal field theory (CFT) living on the asymptotic boundary. This duality is best understood in a regime where the gravity theory is classical and the dual field theory is strongly coupling. It is a well-known fact that the stress-energy tensor in CFT is traceless: $\langle T^\mu_{\;\;\mu}\rangle=0$. This property holds both at the classical level and at the quantum level on flat spacetime, but when the CFT is coupled to a curved background, a Weyl anomaly appears. In the case of a $2d$ $\mathrm{CFT}$, we have  $\langle T^\mu_{\;\;\mu}\rangle=\frac{c}{24\pi}R$, where $R$ is the Ricci scalar, and $c$ is the central charge. Using the holographic dictionary and holographic renormalization, one can compute the expectation value of the holographic stress-energy tensor purely from the classical gravity side and obtain the result $\langle T^\mu_{\;\;\mu}\rangle=\frac{l}{16\pi G}R^{(0)}$, where $R^{0}$ is Ricci scalar obtained from the boundary metric $g^0$, that can be read from the asymptotic expansion (in Fefferman-Graham gauge) of the bulk metric \cite{Henningson:1998gx}. Comparing the last two formulas, we obtain a holographic prediction: $3d$ gravity is dual to a $2d$ CFT with central charge given by $c=\frac{3l}{2G}$. We refer the interested reader to \cite{Ciambelli:2022vot}, where one can find a complete set of references on the topic and this example is proposed as a solved exercise.

\subsection{The Bondi-van der Burg-Metzner-Sachs (BMS) asymptotic symmetries} \label{bms4}
Time has come to study a gravitational example in detail. We will consider an asymptotically flat spacetime in four dimensions and throughout this section we will work with $8\pi G=1=c$ conventions with metric signature being $(-,+,+,+)$. Referring to the previous section, we will follow the steps (i)-(iii) to define our classical theory hence we will specify the dynamics, the boundary conditions and the gauge fixing. Useful references for this section are \cite{Bondi:1960jsa, McCarthy, Barnich:2009se, Barnich:2011mi, Strominger:2014pwa,  Alessio:2017lps, Freidel:2021fxf}.

\subsubsection{Dynamics} 
The bulk dynamics of the theory under exam is defined via the following Einstein-Hilbert action
\begin{equation}
    S_{EH}=\frac{1}{2}\int R \sqrt{\abs{\det{g}}} \dd[4]{x}\;,
\end{equation}
with zero cosmological constant. The goal of this section is to extrapolate the asymptotic charges at the future null infinity of the spacetime. We will show that the symmetries in the bulk are those defined by the Poincaré group which are finite-dimensional, while those defined on the boundary are infinite-dimensional and form the so-called $\mathrm{BMS}$ group. The set of coordinates we will use is $(u,r,\sigma^A)$, where
\begin{itemize}
    \item $u$ is the null time that follows outgoing null geodesic congruences without vorticity,
    \item $r$ is the radial coordinate,
    \item $\sigma^A$ are spatial coordinates on the codimension-2 surface (they can be thought of as the angular coordinates $(\theta,\phi)$ on the sphere).
\end{itemize}
In this set of coordinates, the bulk metric is written as
\begin{equation}
    \dd s^2=-2 e^{2\beta} \dd u \left(\dd r+ F \dd u\right)+r^2 q_{AB} \left(\dd\sigma^A-U^A\dd u\right)\left(\dd\sigma^B-U^B\dd u\right)\;,
\end{equation}
where $q_{AB}$ is the codimension-2 metric and generally $\beta, F, U^A, q_{AB}$ are functions of the full coordinates $(u,r,\sigma^A)$ defined in the bulk. 

\subsubsection{Boundary conditions} 

We assume the following falloffs:
\begin{align}\label{f1}
    g_{ur}&=-1+\order{\frac{1}{r^2}}\;,\\ \label{f2}
    g_{uA}&=\order{1}\;,\\ \label{f3}
    g_{uu}&=\order{1}\;,\\ \label{f4}
    q_{AB}&=\order{1}\;.
\end{align}
Given these falloffs, the causal structure of the spacetime is reported in the figure below, \cref{fig:CD1}. 

\begin{figure}[hbt!]
\begin{center}
\includegraphics[scale=0.4]{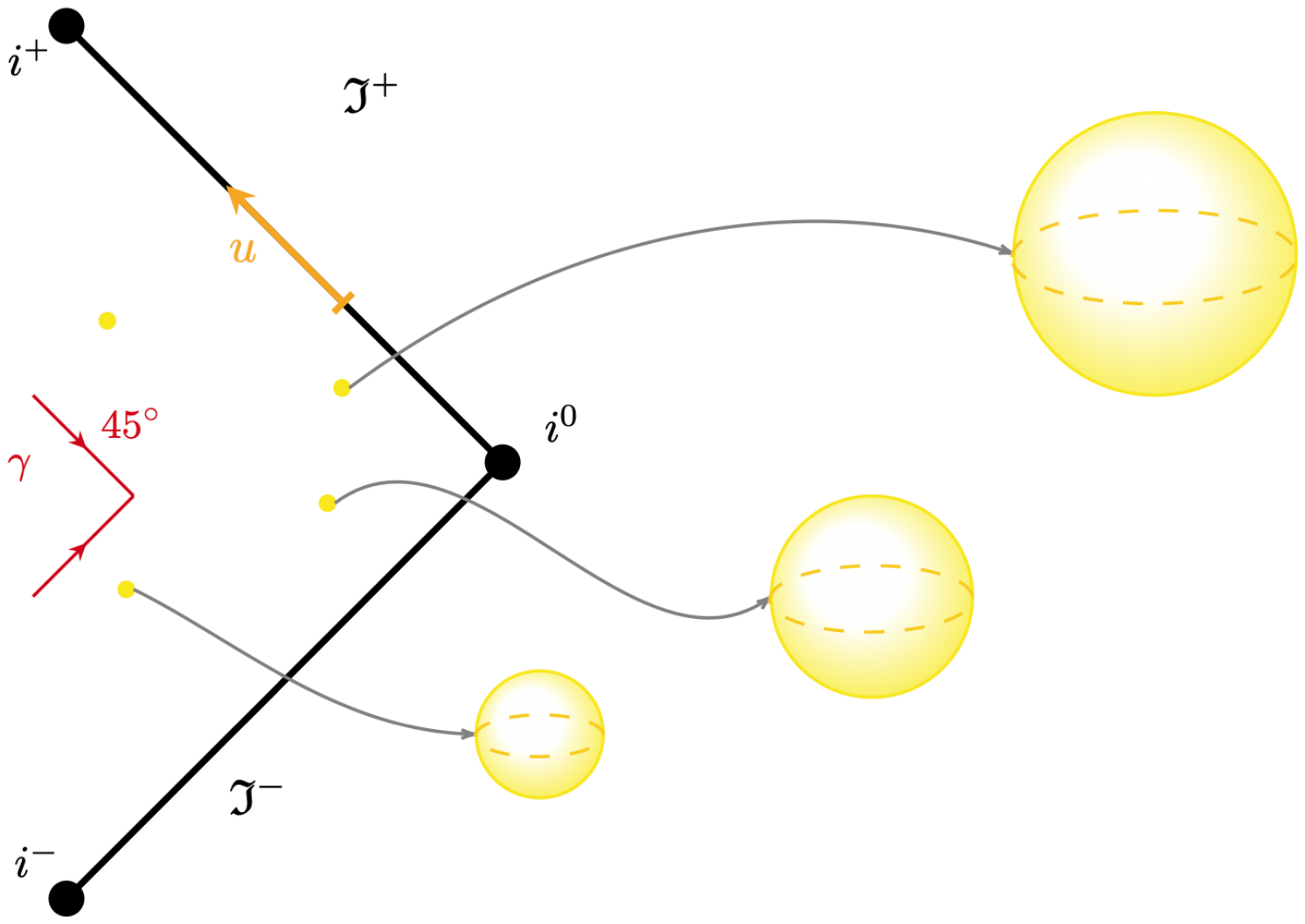}\caption{Conformal diagram of the bulk.}  \label{fig:CD1}
\end{center}
\end{figure}

Firstly, the two black lines drawn at $45^\circ$ represent the future null infinity and past null infinity, labelled with $\mathfrak{I}^+$ and $\mathfrak{I}^-$ respectively. The points $i^0$, $i^-$ and $i^+$ are spacelike infinity, timelike past infinity and timelike future infinity respectively. The $u$-direction is parallel to the $\mathfrak{I}^+$-line while the $r$-direction is radial. Each point in this diagram is of codimension-2 which means that we can think of its structure as that of a sphere. Therefore, the point following the radial direction can be pictured as a \emph{sphere} having increasing radius as the point approaches null infinity. Since a photon is a particle with zero mass, it moves on null geodesics by definition, so it moves following straight $45^\circ$-lines, represented by red lines in the figure. In terms of a photon path, $\mathfrak{I}^-$ represents the initial surface from which photons propagate into the bulk while $\mathfrak{I}^+$ is the final surface reached asymptotically by photons.\\
The bulk dynamics can be, for instance, that of a star forming a black hole. Referring to \cref{fig:CD2}, at $\mathfrak{i}^-$ there is the star, our massive object, which collapses into a black hole whose horizon is represented by the blue line. during its collapse, the radiation emitted can be detected in the vicinity of $\mathfrak{I}^+$. A practical example of this dynamics is given by two infalling spiraling black holes whose radiation emitted in form of gravitational waves is detected at (near) future null infinity (see \cref{fig:CD2}).

\begin{figure}[hbt!]
\begin{center}
\includegraphics[scale=0.5]{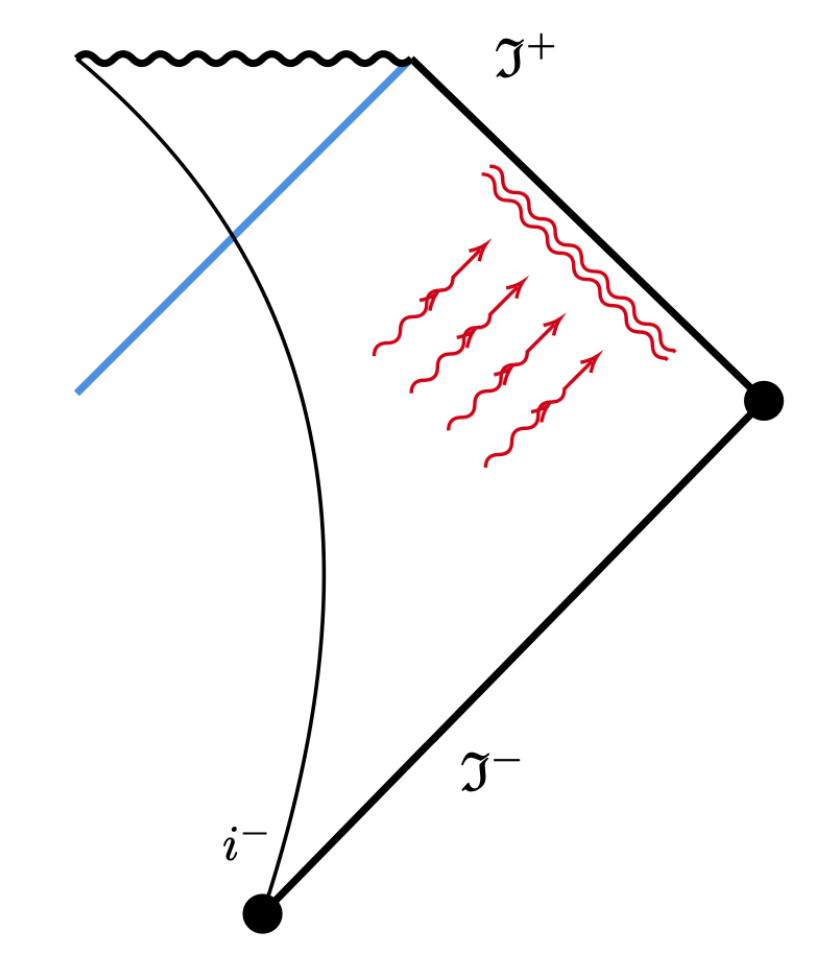}\caption{Conformal diagram of a collapsing star.}  \label{fig:CD2}
\end{center}
\end{figure}

\subsubsection{Gauge fixing} We consider the so-called \emph{Bondi gauge} which consists in enforcing the following constraints ($\sqrt{q}=\sqrt{\det{q}}$)
\begin{equation}\label{gauge}
    g_{rr}=0\ , \quad g_{rA}=0\ , \quad \partial_r
\sqrt{q}=0\;,
\end{equation}
along with the following asymptotic expansions of the bulk fields
\begin{align}\label{totalmass}
    F(u,r,\sigma^A)&= \Bar{F}(u,\sigma^A)-\frac{M}{r}+\dots\;,\\
    \beta(u,r,\sigma^A)&= \frac{\Bar{\beta}(u,\sigma^A)}{r^2}+\dots\;,\\
    q_{AB}(u,r,\sigma^A)&= \Bar{q}_{AB}(u,\sigma^A)+\frac{c_{AB}}{r}+\dots\;,\\ \label{totmom}
    U^A(u,r,\sigma^A)&= \frac{\Bar{U}^A(u,\sigma^A)}{r^2}-\frac{2}{3 r^3}\Bar{p}^A+\dots\;,
\end{align}
where the dots stand for subleading terms. The important thing to note is that the $r$ dependency on the right hand side (RHS) of the previous expressions is explicit. This means that the Bondi gauge completely decouples the radial dependence. Additionally, $M$ will turn out to be (as we will later show) the mass of the system that one would observe while $\Bar{p}$ is its total angular momentum. Moreover, $c_{AB}$ is called the \emph{shear} tensor and carries the gravitational waves information of the system. This object is in fact a $2\times 2$ traceless, symmetric matrix in terms of which we define the \emph{news tensor} $N_{AB}$
\begin{equation}\label{news}
    N_{AB}=\partial_u c_{AB}\ .
\end{equation}

Now that we have defined the theory we move on to the analysis of symmetries which will lead us to the computation of  charges.  

\subsubsection{EOM} 

Let us start by writing down the equations of motion (EOM) obtained by solving Einstein's equations order by order as $r$ goes to infinity. 

They can be divided into two types of EOMs. Firstly, we have four constraints which read
\begin{eqnarray}
    &&\partial_u \Bar{q}_{AB}=0\;,\\
    &&\Bar{\beta}+\frac{1}{32} c_{AB} c^{AB}=0\;,\\
    &&\Bar{R}-4\Bar{F}=0\;,\\
    &&\Bar{U}^{A}+\frac{1}{2}\Bar{D}_B c^{AB}=0\;,
\end{eqnarray}
where $\Bar{R}$ is the codimension-2 Ricci scalar of the boundary metric $\Bar{q}_{AB}$ and $\Bar{D}$ is the covariant derivative with respect to the metric $\Bar{q}_{AB}$. 

The other type are evolution equations since they describe the temporal evolution of the system. For the sake of simplicity we focus our attention only on the \emph{Bondi mass-loss formula}
\begin{equation}\label{bondiformula}
    \partial_u M=-\frac{1}{8} N_{AB} N^{AB} +\frac{1}{4}\Bar{D}_A \Bar{D}_B N^{AB}\;.
\end{equation}
This is called “mass-loss formula'' since, as we will shortly show, $M$ is indeed the charge associated to $\partial_u$, namely to time translations, and is therefore the energy of the system. The equation above also helps us in understanding why $N_{AB}$ is called “news tensor'': it is basically providing us information about the source by telling us how the latter affects the mass dynamics.\\
This EOM can be integrated on a sphere $S$ leading to
\begin{equation}
    \partial_u \int_S \sqrt{\Bar{q}} M=-\frac{1}{8} \int_S \sqrt{\Bar{q}} N_{AB}N^{AB}\;,
\end{equation}
where the covariant derivative term is dropped out since it is a total derivative contribution on the sphere. Note that if the RHS is null then the mass of the system is conserved while, on the contrary, its time evolution is negative. This means that, given a finite mass at initial time, as time evolves radiation is emitted resulting in a mass-decreasing mechanism. This is exactly the reason why the Bondi equation is known as the \emph{mass-loss formula}.\\
These are the EOMs that we will analyze later on. Note that of course there are subleading EOMs, however we are only interested in the leading  asymptotic ones, since we will exclusively deal with the field space asymptotically to the boundary.

\subsubsection{Symmetries} 

By construction, a symmetry \emph{must} preserve the chosen gauge. Therefore,  we impose that the Bondi gauge in \eqref{gauge} is unaffected under diffeomorphisms, namely when acting with the Lie derivative. Hence, we impose the following
\begin{equation}
    \mathcal{L}_\xi g_{rr}=0\ , \quad \mathcal{L}_\xi g_{rA}=0\ , \qq{and} \mathcal{L}_\xi \partial_r (\sqrt{q})=0\;, 
\end{equation}
 which can be solved leading to
 \begin{align}
     \xi^u&= \tau(u,\sigma)\;,\\
     \xi^A&= Y^A(u,\sigma)-I^{AB}\partial_B \tau(u,\sigma)\;,\\
     \xi^r&=-r w(u,\sigma)+\frac{r}{2}\left[D_A\left(I^{AB}\partial_B \tau(u,\sigma)+U^A\partial_A\tau(u,\sigma)\right)\right]\;,
 \end{align}
 where 
 \begin{equation}
     I^{AB}=\int_r^\infty \frac{\dd r^\prime}{r^{\prime\ 2}}\ e^{2\beta}\ q^{AB}\;,
 \end{equation}
and $T(\sigma)$, $Y^A(\sigma)$ and $w(\sigma)$ are arbitrary functions parameterizing $\xi$.
 The next step is to impose the falloffs as well. Again, looking at \eqref{f1}-\eqref{f4} we immediately have that
 \begin{align}
     \mathcal{L}_\xi g_{ur}&=\order{\frac{1}{r^2}}\;,\\
     \mathcal{L}_\xi g_{uu}&=\order{1}\;,\\
     \mathcal{L}_\xi g_{uA}&=\order{1}\;,\\
     \mathcal{L}_\xi g_{AB}&=\order{r^2}\;,
 \end{align}
 which result in constraining the functions $Y^A(u,\sigma)$, $w(u,\sigma)$ and $\tau(u,\sigma)$ to be
 \begin{align}
     Y^A(u,\sigma)&=Y^A(\sigma)\;,\\
     w(u,\sigma)&=w(\sigma)\;,\\
     \tau(u,\sigma)&=T(\sigma)+u w(\sigma)\;,
 \end{align}
 and, as a result, the vector field $\xi$ at first leading order in $r$ becomes
\begin{equation}\label{vectorxi}
     \xi= T(\sigma)\partial_u+ Y^A(\sigma) \partial_A+w(\sigma)(u\partial_u-r\partial_r)+\dots\ .
\end{equation}
This vector field will give rise to conserved quantities. Moreover, since $T(\sigma)$, $Y^A(\sigma)\partial_A$ and $w(\sigma)$ parameterize $\xi$, they are actually the parameters of the symmetries coming from the residual diffeomorphisms. Having obtained $\xi$ we can move on to write down the residual symmetry algebra.

\subsubsection{Algebra}

Using \eqref{vectorxi} the algebra of symmetries of the theory is
\begin{equation}
     \comm{\xi_1}{\zeta_2}= T_{12}\partial_u+ Y^A_{12}\partial_A+w_{12}(u\partial_u-r\partial_r)\;,
\end{equation}
where
\begin{align}
     &Y^A_{12}\defeq\comm{Y_1}{Y_2}^A\;,\\
     &w_{12} \defeq Y_1(w_2)-Y_2(w_1)\;,\\
     &T_{12} \defeq Y_1(T_2)-w_1(T_2)-Y_2(T_1)+w_2(T_1)\;.
\end{align}
Since we have not computed the charges yet, these symmetries are residual symmetries. Therefore they can be trivial ones or physical ones. We will determine it in a moment by computing the associated charges. 
 
Let us now make some important observations. We know that the conformal group on the $2$-sphere is $SO(1,3)$ which generalises to $SO(1,d+1)$ when considering the $d$-sphere. As a result, if $Y^A\partial_A$ were conformal Killing vectors on the $2$-sphere, the group of symmetries that we would have obtained from it is $SO(1,3)$, which is indeed the Lorenz group. Additionally, $\xi$ has the $T(\sigma)\partial_u$ contribution which is exactly the one defining the so-called supertranslations. Therefore, focusing for a moment only on the $T(\sigma)$ and $Y^A(\sigma)$ terms in $\xi$, we see an enhancement of the Poincaré group. It is an enhancement because these parameters are a priori arbitrary functions on the $2$-sphere hence they are infinitely many. If we were restricted to globally well-defined conformal Killing vectors on the $2$-sphere, and constant translations, we would get the usual finite Poincaré group $SO(1,3) \ltimes \mathbb{R}^4$. Nevertheless, what we obtained is an enhancement of it,  namely the semi-direct product of diffeomorphisms of the $2$-dimensional space and supertranslations.\\
To have a visual representation and intuition of the supertranslation $T(\sigma)\partial_u$ (see \cref{fig:CD3}), just imagine a past light cone of an observer $O$. A cut of this cone is the surface $S$ and a supertranslation is an angle-dependent translation of this surface along the $u$-direction, allowing its points to move with different velocities. 

\begin{figure}[hbt!]
\begin{center}
\includegraphics[scale=0.5]{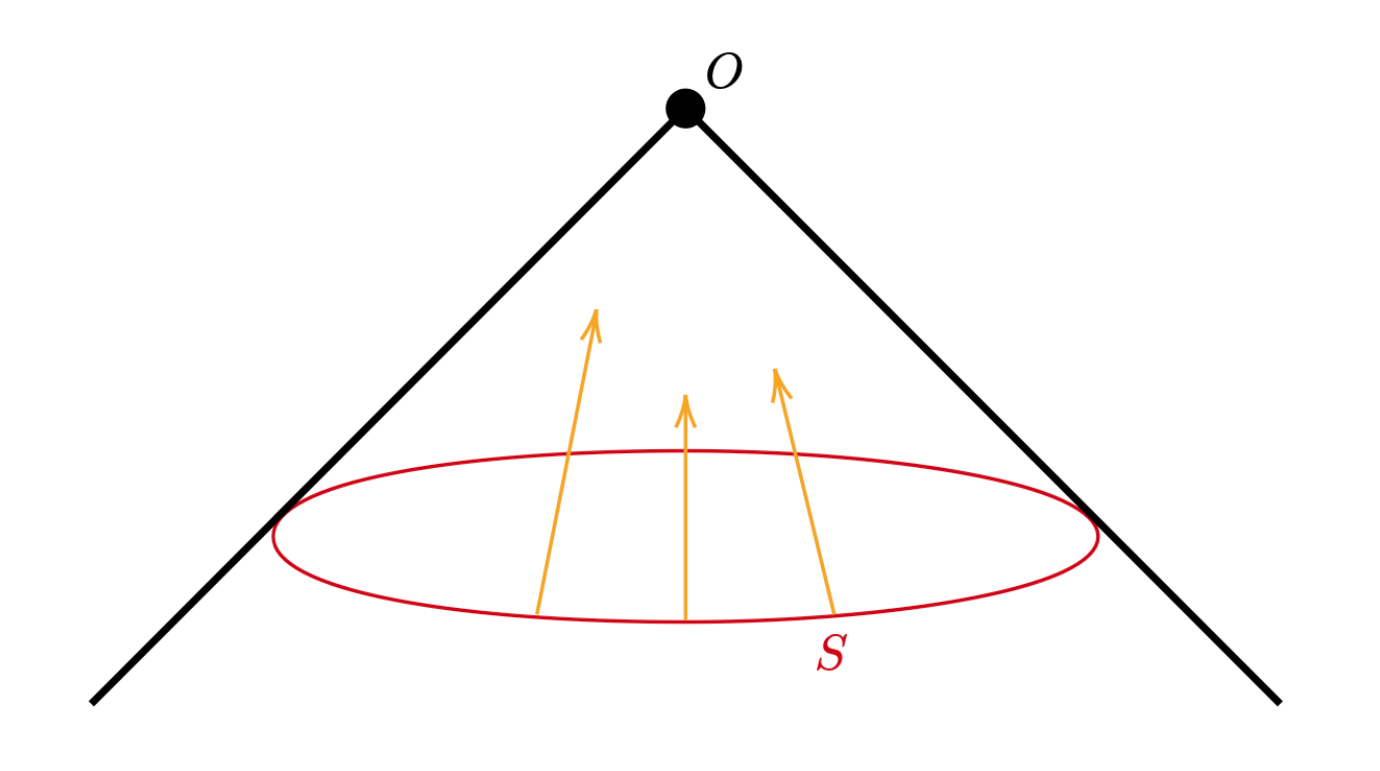}\caption{Sketch of the past light cone of an observer $O$ representing the action of a supertranslation.}  \label{fig:CD3}
\end{center}
\end{figure}

 The algebra we obtained is known as the \emph{generalised BMS algebra} \cite{Campiglia:2014yka}
 \begin{equation}
    \text{generalised BMS algebra}\; =\;  \mathrm{diff}(S) \loplus \mathbb{R}^S\;.
 \end{equation}
In our treatment, we also had another generator, called $w(\sigma)$, which enhances the algebra to the Weyl BMS algebra studied in \cite{Freidel:2021fxf}, denoted as BMSW algebra\footnote{This algebra includes supertranslations, local Weyl rescalings and arbitrary
diffeomorphisms of the $2$-dimensional sphere metric.}. 
We are ready to finally compute the charges of the theory. We will show that the $\mathrm{BMSW}$ algebra is indeed charged, and thus its generators are asymptotic symmetries.

 \subsubsection{Charges} \label{charges} Let us recall that at leading order in $r$, the metric we are considering contains $M, \Bar{\beta},\Bar{U}^A, c_{AB}$ as parameters and the equation to solve to obtain the charges is
 \begin{equation}
     \dd Q_{\hat{\xi}}=I_{\widehat{\xi}}\, \theta_{EH}-i_\xi L_{EH}\;.
 \end{equation}
 As a consequence, this means that we need to know how our fields transform under the symmetries. This latter task is solved by computing the Lie derivative of the metric at leading orders,
 so that we can extrapolate $\cL_\xi M$, etc.\\
 Having setting the stage, we complete these preliminary information by recalling that $ L=\frac{1}{2} R \sqrt{|\det{g} | }\ \dd[4]{x}$ and using $\delta L=G_{\mu\nu}\delta g^{\mu\nu}+\dd \theta$ we have
 \begin{equation}\label{thetaEH}
     \theta_{EH}= \frac{1}{2}\left(g^{\mu \nu}\delta \Gamma^\alpha_{\mu\nu}-g^{\alpha\mu}\delta\Gamma^\nu_{\mu\nu}\right)\epsilon_\alpha
 \end{equation}
 with $\epsilon_\alpha$ being the codimension-1 volume form.\\ 
 We skip all the technicalities and give the final expressions of the charge associated to $T$, $w$ and $Y$, respectively. For each of the charges associated to a parameter, all the others are set to zero. In order:
 \begin{itemize}
     \item \textbf{$H_{\hat{T}}$}\\
     Setting $Y=0=w$ we have that the charge associated to supertranslations reads
     \begin{equation}\label{HTu}
         H_{\hat{T}}=\int_S Q_{\hat{\xi}}=\int_S \sqrt{\Bar{q}}\, T(\sigma) \left(M-\frac{1}{2}\Bar{D}_A\Bar{U}^A\right)\;,
     \end{equation}
     from which it immediately follows that if $T$ is $\sigma$-independent, the last term on the RHS would drop since it is a total derivative contribution and the charge would be the total mass of the system. We have thus shown why we were so confident in calling $M$ in eq. \eqref{totalmass} the total mass of the system. It is important to stress here that this result is finite, there is no need for renormalization. On the contrary, the other two charges turn out to be divergent and renormalization is needed. For the sake of simplicity we only report the finite part, which is the endpoint of the renormalization procedure, and refer to \cite{Freidel:2021fxf} for more details.
     
     \item \textbf{$H_{\hat{w}}$}\\
      After setting $T=0=Y$ the finite part is
     \begin{equation}
    H_{\hat{w}}^{\text{finite}}=\int_S\sqrt{\Bar{q}}\, w(\sigma)\left[4\Bar{\beta}+u\left(M-\frac{1}{2}\Bar{D}_A\Bar{U}^A\right)\right]\;.
     \end{equation}
     
      \item \textbf{$H_{\hat{Y}}$}\\
    After setting $T=0=w$, again the finite part reads 
    \begin{equation}
        H_{\hat{Y}}^{\text{finite}}=\int_S \sqrt{\Bar{q}}\, Y^A(\sigma) \left(\Bar{p}_A+2 \Bar{D}_A \bar{\beta}\right)\;,
    \end{equation}
    which, as for the charge associated to supertranslations, immediately tells us that if $Y^A$ is $\sigma$-independent the last term on the RHS drops and the charge associated to the diffeomorphisms on the $2$-dimensional space becomes the total angular momentum. As before, this is the reason why in eq. \eqref{totmom} we called $\Bar{p}$ the total angular momentum of the system.
 \end{itemize}
 Let us pause for a moment to appreciate that we have just obtained the observables of gravity and we have demonstrated that they arise as Noether (corner) charges associated to diffeomorphisms.
 
 \subsubsection{Charge algebra} 
 
This will be nothing else than the representation of the $\mathrm{BMSW}$ algebra. In the presence of symplectic fluxes as we already discussed  the final result is \eqref{modiBR}, that is,
 \begin{equation}\label{eomfromsymm}
    \poissonbracket{H_{\hat{\xi}}}{H_{\hat{\zeta}}}=\fL_{\hat{\zeta}} H_{\hat{\xi}}-I_{\hat{\zeta}} F_{\hat{\xi}} +\int_S i_\xi i_\zeta L\;.
 \end{equation}
 We are going to show that the previous expression contains the EOMs, so the evolution of the system expressed via the Bondi mass loss formula emerges from symmetries. To this aim, we consider two supertranslation generators, since their Poisson brackets is zero by definition because the algebra of (super)translations is abelian. We thus choose as generators the following two vectors
 \begin{equation}
     \xi=T\partial_u \qq{and} \zeta=\partial_u\;.
 \end{equation}
 Note that $\zeta$ is the generator of the symmetries associated to the energy of the system. Along with the Poisson brackets, also the last term on the RHS of \eqref{eomfromsymm} is zero when considering two supertranslations because the two contractions are along the same direction. Consequently, the final expression is
 \begin{equation}\label{Handflux}
     \fL_{\hat{\zeta}} H_{\hat{\xi}}=I_{\hat{\zeta}} F_{\hat{\xi}} \quad \implies \quad \fL_{\hat{\partial_u}} H_{\hat{T\partial_u}}=I_{\hat{\partial_u}} F_{\hat{T\partial_u}}\;,
 \end{equation}
 which, using the expression in \eqref{HTu} and in \eqref{thetaEH}, reads
 \begin{equation}\label{bondiformula2}
     \int_S \sqrt{\Bar{q}} T(\sigma) \left(\Dot{M}+\frac{1}{8}N^{AB}N_{AB}-\frac{1}{4}\Bar{D}_A \Bar{D}_B N^{AB}\right)=0\ ,
 \end{equation}
 which is exactly the Bondi mass-loss formula. 

\subsection{The power of symmetries: memory effects}\label{memory sec}
In this section we will introduce the concept of \emph{memory effects} in gravity. We again refer to \cite{Ciambelli:2022vot} for original references on the topic, while this section is mainly based on \cite{Satishchandran:2019pyc}, see also \cite{Strominger:2014pwa}. \emph{Memory effects} stands for permanent relative displacement due to a burst of gravitational waves. More precisely, if the system under consideration consists of a set of massive objects, they will be subject to a permanent displacement after a gravitational wave passes through. In this sense, the masses have memory of the passage of gravitational waves. This effect can be computed from the geodesic deviation equation in general relativity. Consider a freely falling particle moving on a worldline. Let $v^a$ denote the vector tangent to the particle worldline and $\xi^b$ the deviation vector, describing how a direction has changed infinitesimally. The geodesic deviation equation reads
 \begin{equation}\label{displace}
     \left(v^a \nabla_a\right)^2 \xi^b=- \tensor{R}{_a_c_d}{^b}\ v^a v^d \xi^c\ .
 \end{equation}
Our goal in this section will be to derive the vector $\xi$: we want to compute the displacement of the particles. To achieve that, let us consider a number of particles near asymptotic infinity such that at initial time $v^a=\delta^a_u$. Therefore, to leading order in $r$, \eqref{displace} gives ($x^a=(r,x^\mu)$)
\begin{equation}\label{displace2}
    \pdv[2]{\xi^\mu}{u}=- \tensor{R}{_{u\alpha u}}{^\mu}\ \xi^\alpha\; .
\end{equation}

Now, it can be shown that, at leading order when $r\to \infty$ the Riemann tensor, made up by a traceless contribution given by the Weyl tensor plus the Ricci scalar and the Ricci tensor counterparts, simply becomes the Weyl tensor. That is: $\tensor{R}{_{u\alpha u}^\mu} \widesim[2]{r\to\infty} \tensor{\Bar{C}}{_{u\alpha u}^\mu}$ at leading order in $r$ (the bar on the Weyl tensor stands for the leading order contribution). Additionally, if $\xi^\alpha_{(0)}$ denotes the initial displacement and if $\xi^\alpha-\xi^\alpha_{(0)}$ goes to zero fast enough as $r\to \infty$, one can safely replace $\xi^\alpha$ with $\xi^\alpha_{(0)}$ on the RHS of \eqref{displace2} so that
\begin{equation}
    \pdv[2]{\xi^\mu}{u}=-\tensor{\Bar{C}}{_{u\alpha u}}{^\mu} \xi^\alpha_{(0)}\;,
\end{equation}
which, after integrating in time twice, becomes
\begin{equation}
    \xi^\mu\rvert^{u=+\infty}_{u=-\infty}=-\int^\infty_{-\infty} \dd u^\prime  \int^\infty_{-\infty} \dd u\ \tensor{\Bar{C}}{_{u\alpha u}}{^\mu} \xi^\alpha_{(0)}\;.
\end{equation}
This can be recasted as follows:
\begin{equation}\label{displace3}
     \xi^\mu\rvert^{u=+\infty}_{u=-\infty}= \tensor{\Delta}{^\mu}{_\alpha}\  \xi^\alpha_{(0)}\ ,
\end{equation}
where we have introduced the \emph{memory tensor} $\tensor{\Delta}{^\mu}{_\alpha}$ as
\begin{equation}
    \tensor{\Delta}{^\mu}{_\alpha}= -\int^\infty_{-\infty} \dd u^\prime  \int^\infty_{-\infty} \dd u\ \tensor{\Bar{C}}{_{u\alpha u}}{^\mu}\;.
\end{equation}
The memory tensor keeps track of the passage of the gravitational waves originated by an unspecified black hole dynamics and the left hand side (LHS) of \eqref{displace3} gives the evolution between the stationary state in the far past and the stationary state in the far future.

From now on we will work in the Bondi gauge previously introduced. In this gauge a straightforward computation leads us to the following crucial result:
 \begin{equation}
     \Bar{C}_{uAuB}=-\frac{1}{2}\partial_u N_{AB}\;,
 \end{equation}
 where $\{A,B\}$ are coordinates on the $2$-sphere  matching the notation of the previous section and $N_{AB}$ is the news tensor. The news tensor introduced in \eqref{news} is such that
 \begin{equation}\label{deltaAB}
     \Delta_{AB}=\frac{1}{2}\int^{+\infty}_{-\infty}\dd u\ N_{AB}\overset{\eqref{news}}{=} \frac{1}{2} c_{AB}\rvert^{u=+\infty}_{u=-\infty}\;,
 \end{equation}
 from which it is now evident why the memory tensor has information about gravitational waves, it is written in terms of the shear tensor.

Now that we have the memory tensor, we can further investigate its relation with the symmetries. Suppose for instance that in the far past $c_{AB}=0$ corresponding to a \emph{stationary era}. This would for example be the case in a configuration in which two black holes are far away. A natural question is then what is the fate of $c_{AB}$ in the far future. The latter can at best be the BMS-transform of its initial value, that is:
\begin{equation}
    c_{AB}\rvert_{u=-\infty}=0\quad \implies \quad c_{AB}\rvert^{u=+\infty}= \mathrm{BMS} (c_{AB}\rvert_{u=-\infty})\; .
\end{equation}
It can be shown that the variation of the shear tensor under a supertranslation gives ($\Bar{D}_{\langle A}\Bar{D}_{B\rangle}$ is the symmetric traceless part of $\Bar{D}_A\Bar{D}_B$):
\begin{equation}
    \fL_{\hat{T}} c_{AB}=T \partial_u c_{AB}-2 \Bar{D}_{\langle A}\Bar{D}_{B\rangle} T\;.
\end{equation}
Applying this result to $c_{AB}\rvert_{u=-\infty}$ with our initial hypothesis of a stationary era we obtain
\begin{equation}
    c_{AB}\rvert^{u=+\infty} = \fL_{\hat{T}} c_{AB}\rvert_{u=-\infty}=T \partial_u c_{AB}\rvert_{u=-\infty}-2 \Bar{D}_{\langle A}\Bar{D}_{B\rangle} T \underbrace{=}_{c_{AB}\rvert_{u=-\infty}=0}-2 \Bar{D}_{\langle A}\Bar{D}_{B\rangle} T\;.
\end{equation}
The memory tensor \eqref{deltaAB} thus becomes
\begin{equation}
    \Delta_{AB}=- \Bar{D}_{\langle A}\Bar{D}_{B\rangle} T\;,
\end{equation}
from which we can express $T$ as a function of $\Delta_{AB}$. This manipulation is essential when considering the Bondi mass loss formula. Indeed, plugging it into $T(\Delta_{AB})$ we are able to have an expression of the memory tensor in terms of measurable quantities. To have more insight, we move to the wave forms and briefly illustrate the effect of this prediction with a sketch. In \cref{fig:lisa} the wave form of the gravitational wave as a function of time is displayed.
\vspace{0.5em}

\begin{figure}[hbt!]
\begin{center}
\includegraphics[scale=0.5]{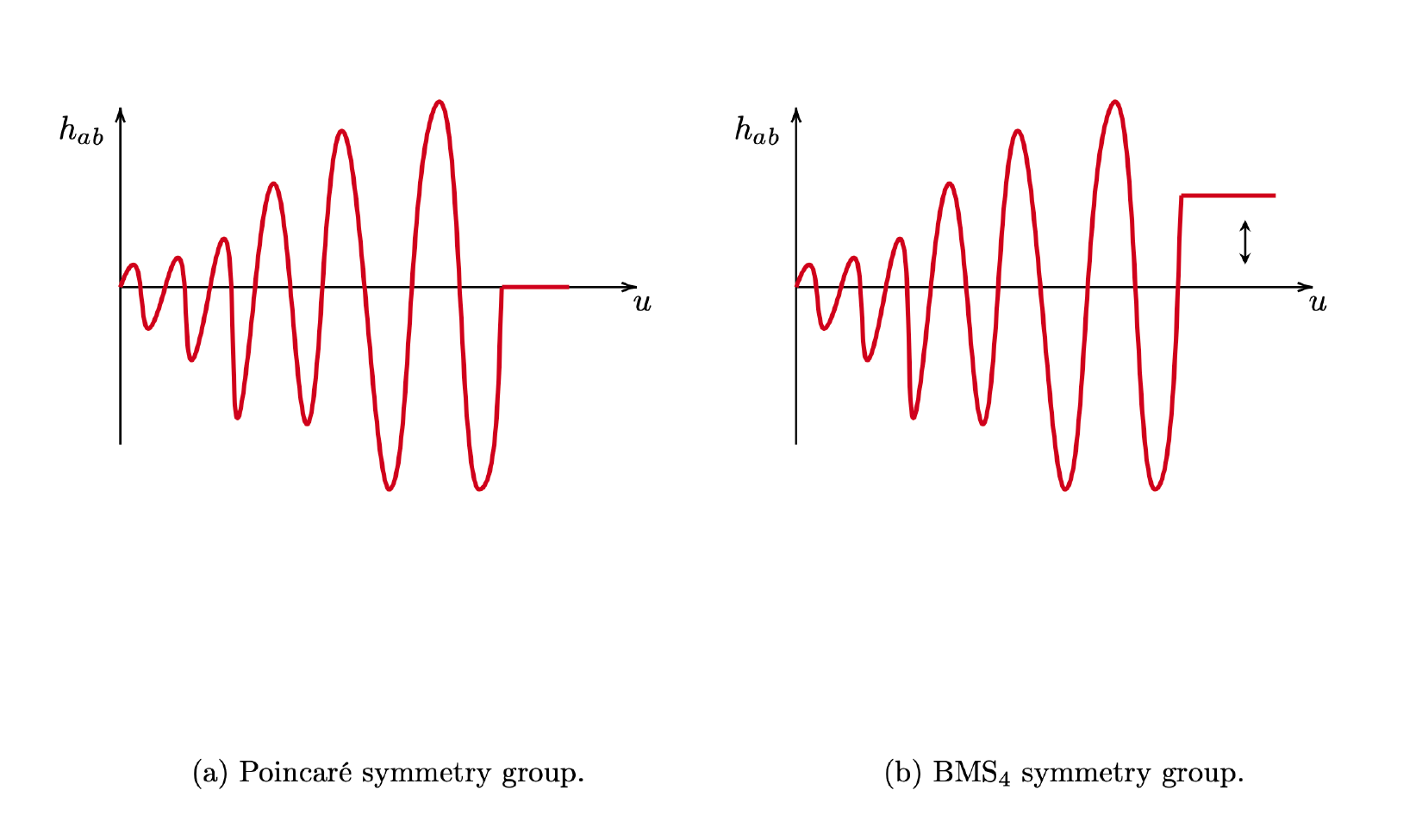}\caption{Sketch of the metric perturbation as a function of time.}  \label{fig:lisa}
\end{center}
\end{figure}

More precisely, on the left of \cref{fig:lisa} the effect is represented with the Poincaré group as a symmetry group while on the right, the group of symmetry is the $\mathrm{BMS}_4$ symmetry group. There is a clear mismatch between the two figures due to the presence of an offset in the second case. Let us explain better what information is hidden in this \cref{fig:lisa}. Firstly, at some time, the profile becomes constant. Suppose two black holes collide and emit gravitational waves, their passage in our detector is highlighted by the oscillating behaviour of the function in the figure. After this passage, however, there is a remnant of the presence of the gravitational waves: the memory displacement. Consequently, this displacement that we have studied in this section culminates in the prediction of a constant shift in \cref{fig:lisa}. Furthermore, we expect that the latter could be detected and measured by upcoming experiments.

We conclude this section with a concept introduced in 2017 by Strominger \cite{Strominger:2017zoo}, the \emph{infrared triangle} that well summarises what we have learned so far. In the triangle depicted in \cref{fig:triangle}, all the vertices are equivalent. Once one vertex is known, the others can be attained by mathematical manipulations. We however remark that we have followed here an approach based entirely on symmetries. We have shown how symmetries are so powerful that memory effects can be derived from them. This is the logic of the \emph{corner proposal}, enunciated in the next section, in which symmetries are exploited to infer new constraints in quantum gravity. Before doing so, we fill the last tip of the triangle and show how symmetries can be used to deduce the infrared soft theorems.

\begin{figure}[hbt!]
\begin{center}
\includegraphics[scale=0.4]{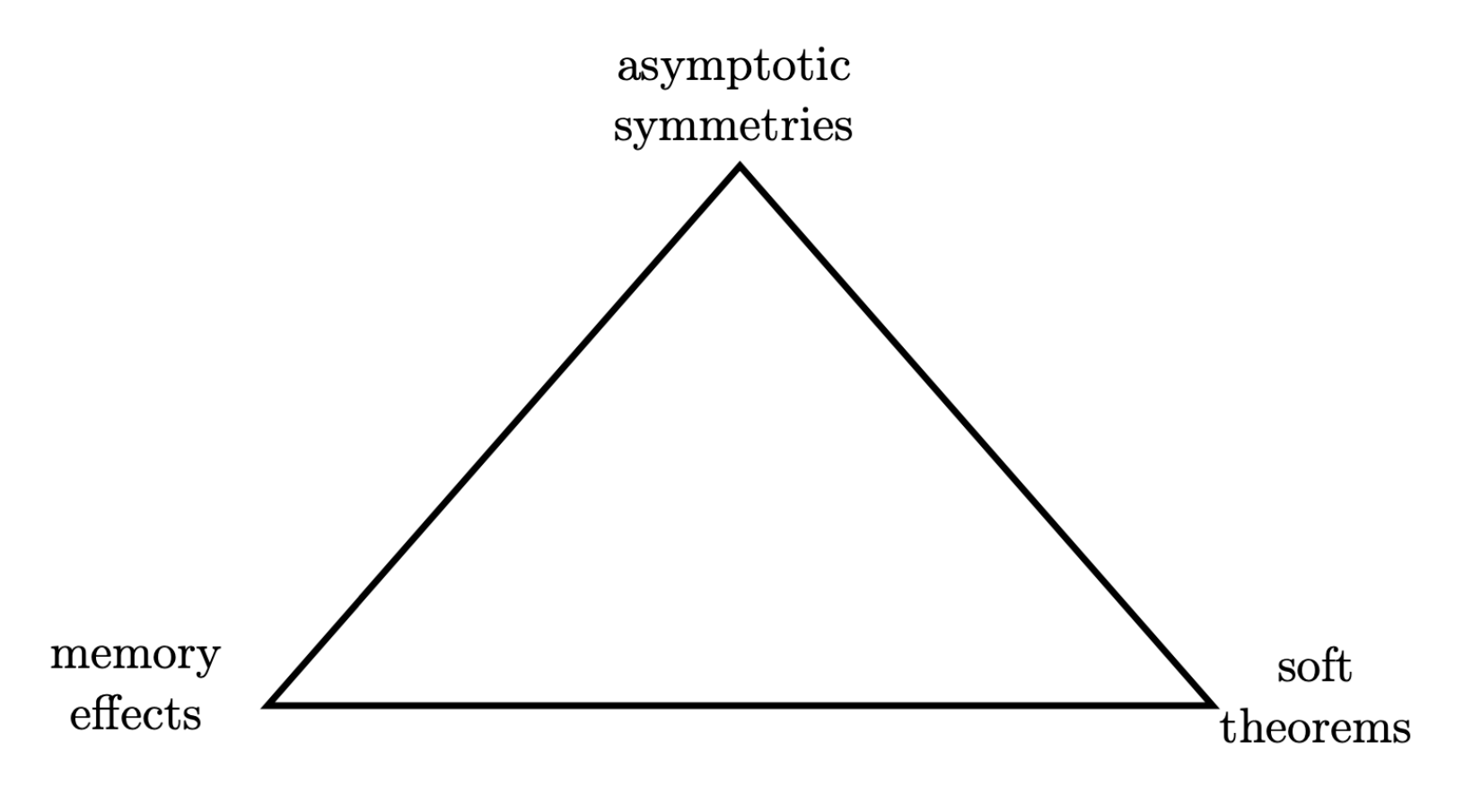}\caption{The infrared triangle.}\label{fig:triangle}
\end{center}
\end{figure}

\subsection{The power of symmetries: soft theorems}\label{soft theorems sec}

In this section we show how symmetries can be used to derive Weinberg's soft theorems \cite{Weinberg:1965nx}. The relationship between these two topics was observed in \cite{He:2014laa}, which we use here as main reference, and we again refer to \cite{Ciambelli:2022vot} for more references. Soft theorems establish a connection between the amplitude of a scattering process involving one soft particle and the one involving only hard particles, where the soft particle contribution factorises. 

We start computing the asymptotic phase space Poisson bracket. In order to do so, one can show that the renormalised symplectic two-form takes the form
\begin{equation}
    \Omega^R=\int_{\mathfrak{I}}\frac{1}{4}\delta N_{AB}\wedge \delta\qty(\sqrt{\Bar{q}}c^{AB})+\dots\; .
\end{equation}
This is the famous Ashtekar-Streubel result \cite{Ashtekar:1981bq}. Using this two-form, one can compute the Poisson bracket between $N_{AB}$ and $c_{AB}$, and obtain
\begin{equation}
    \poissonbracket{N_{AB}(u,z,\Bar{z})}{c_{CD}(u',w,\Bar{w})}=2\gamma_{<AC}\gamma_{BD>}\delta(u-u')\delta^2(z-w)\;,
\end{equation}
where $z$ and $\Bar{z}$ are coordinates on the sphere $S^2$.
As a side note, we mention that Ashtekar promoted this Poisson bracket to a commutator, using standard rules of canonical quantization ($\{,\}\rightarrow \frac{1}{i\hbar}[,]$), therefore quantizing gravity without requiring the Newton constant to be small. This is possible only asymptotically, where this quantization is still made on a classical geometric background.

Let us now define the Fourier transformed field $c_{AB}(\omega,z,\Bar{z})$, for $\omega\geq 0$, as 
\begin{equation}
    c_{AB}(\omega,z,\Bar{z})=\int_{-\infty}^{+\infty}c_{AB}e^{i\omega u}\mathrm{d}u \; .
\end{equation}
We can then define
\begin{align}
    a_{+}(\omega,z,\Bar{z})&\equiv 2\pi i c_{zz}(\omega,z,\Bar{z})\;,\\
     a_{-}(\omega,z,\Bar{z})&\equiv 2\pi i c_{\Bar{z}\Bar{z}}(\omega,z,\Bar{z})\;, 
\end{align}
such that $a_{\pm}^{*}$ become positive/negative-helicity graviton creation operators. The field $c_{AB}$ is part of the metric, and therefore we can use this field, previously shown to be related to gravitational waves, to create/annihilate gravitons. Furthermore, we can compute
\begin{equation}
    \poissonbracket{a_{\pm}(\omega,z,\Bar{z})}{a_{\pm}^{*}(\omega',w,\Bar{w})}=-2i\omega(2\pi)^3\delta(\omega-\omega')\delta^2(z-w)\;,
\end{equation}
which proves that $a_{\pm}$ satisfy the canonical creation/annihilation operators algebra.

For the rest of this section, we will have to slightly modify the definition of charges and fluxes. Calling $T$ the supertranslation generator and $M$ the mass aspect, we define\footnote{For simplicity we assume that the space is the projective plane, with metric $\dd s^2=2\dd z \dd \bar z$, such that $\sqrt{\Bar{q}}=1$.}
\begin{equation}
    \Tilde{H}_{\hat{T}}=2\int_S TM \;,
\end{equation}
together with an appropriate expression for the flux. It can be proved that in this case we have
\begin{equation}\label{chargesm}
    \Tilde{H}_{\hat{T}}^{+}-  \Tilde{H}_{\hat{T}}^{-}=-\int_{\mathfrak{I}^+} T {\cal F}+\frac{1}{2}\int_S T\Bar{D}_A\Bar{D}_B
\Delta^{AB} \; ,
 \end{equation}
where we introduced the local Bondi hard flux, defined as
\begin{equation}
    {\cal F}=\frac14 N_{AB} N^{AB}\;.
\end{equation}
For simplicity, we use the notation  ${H}_T^{+}\equiv {H}_T(u=+\infty)$. We have shown previously that $\Delta_{AB}=c_{AB}^+-c_{AB}^-$. We would like to interpret the first term as the hard charge and the second term as the soft charge. To see this, note that we have 
\begin{align}
  c_{AB}^+-c_{AB}^-=\int_{-\infty}^{+\infty}\partial_u c_{AB}=\lim _{\omega\rightarrow 0}\int_{-\infty}^{+\infty} e^{i\omega u }\partial_u c_{AB}=\lim _{\omega\rightarrow 0}\int_{-\infty}^{+\infty} -i\omega c_{AB}(\omega,z,\Bar{z})\;,
\end{align}
where we integrated by parts and used $\lim_{ \omega\rightarrow 0} (e^{i\omega u}c_{AB})^+_{-}=0$.
We also impose one extra condition
\begin{equation}
   {\bar D}_z{\bar D}_z(c_+^{zz}-c_{-}^{zz})= {\bar D}_{\Bar{z}}{\bar D}_{\Bar{z}}(c_+^{\Bar{z}\Bar{z}}-c_{-}^{\Bar{z}\Bar{z}})\;,
\end{equation}
satisfied by the so called \textit{Christodoulou-Klainerman spacetimes} \cite{SEDP_1989-1990____A15_0,Strominger:2013jfa}. Using this equation, we rewrite the second term in (\ref{chargesm}) as 
\begin{equation}
  Q_{\mathrm{soft}}= \frac{1}{2}\int_S T\Bar{D}_A\Bar{D}_B
\Delta^{AB}= \frac{1}{2\pi} \lim _{\omega\rightarrow 0} \omega \int_S T {\Bar{D}}_{\Bar{z}}{\Bar{ D}}_{\Bar{z}}a_{+}(\omega,z,\Bar{z})\;.
\end{equation}
For this reason we call this term a soft charge (word soft originating from the limit $\omega\rightarrow 0$) and denote it $Q_{\mathrm{soft}}$. This term corresponds to a conformally soft graviton insertion.
Analogously, the first term in  (\ref{chargesm}) is called the hard charge, denoted  $Q_{\mathrm{hard}}$. 

For simplicity, we consider asymptotic $\ket{in}$ and $\ket{out}$ states defined as a cloud of massless particles. Charge conservation implies
\begin{equation}\nonumber
     \Tilde{H}_{\hat{T}}^{+}-  \Tilde{H}_{\hat{T}}^{-}|_{\mathfrak{I}^+}= \Tilde{H}_{\hat{T}}^{+}-  \Tilde{H}_{\hat{T}}^{-}|_{\mathfrak{I}^-}\;.
\end{equation}
At the quantum level, charge conservation is expressed as the fact that the charge operator commutes with the $S$ matrix, that is, $\bra{out}[Q,S]\ket{in}=0$. This expression is a Ward Identity for a given symmetry. We thus have
\begin{equation}
    Q_{\mathrm{hard}}+Q_{\mathrm{soft}}|_{\mathfrak{I}^+}=Q_{\mathrm{hard}}+Q_{\mathrm{soft}}|_{\mathfrak{I}^-}\Rightarrow\bra{out}[Q_{\mathrm{soft}},S]\ket{in}+\bra{out}[Q_{\mathrm{hard}},S]\ket{in}=0\;.
\end{equation}
The last equation is a Ward identity for BMS, where here we only used supertranslations.
Now, since $Q_{\mathrm{hard}}$ is the energy operator for incoming particles, we have
\begin{equation}
   Q_{\mathrm{hard}}\ket{in}=\sum_k E_k T(z_k,\Bar{z}_{k})\ket{in},
\end{equation}
together with a similar expression for $out$. In our case on the other hand we have
\begin{equation}
    Q_{\mathrm{soft}}\ket{in}\sim a_{+}\ket{in}=0\;.
\end{equation}
Using this, we obtain
\begin{equation}
    \frac{1}{2\pi}\lim_{\omega_g\rightarrow 0}\int_S \mathrm{d}^2 w T(w,\Bar{w}){\bar D}_{\Bar{w}}{\bar D}_{\Bar{w}}\bra{out }a_+(\omega_g,w,\Bar{w})S\ket{in}=-\sum_k E_k T(z_k,\Bar{z}_k)\bra{out}S\ket{in},
\end{equation}
where $\omega_g$ is the soft graviton frequency.
This equation is true for any supertranslation $T$. We can thus apply it to the special supertranslation
\begin{equation}
    T(z,\Bar{z})=\frac{\Bar{w}-\Bar{z}}{w-z}\; .
\end{equation}
Using the following identity: 
\begin{equation}
{\bar D}_{\Bar{w}}{\bar D}_{\Bar{w}}\Big(\frac{\Bar{w}-\Bar{z}}{w-z}  \Big)=2\pi \delta^2(w-z)\;,
\end{equation}
we can finally write down
\begin{equation}
    \lim_{\omega_g\rightarrow 0} \omega_g\bra{out}a_+(\omega_g,w,\Bar{w})S\ket{in}=-\sum_k E_k \frac{\Bar{w}-\Bar{z}_k}{w-z_k}\bra{out}S\ket{in}.
\end{equation}
This is the desired identity, expressing Weinberg's soft graviton theorem in coordinates representation. It is often stated in  momentum representation, where it has the form
\begin{equation}\label{eq1}
    -\sum_k E_k \frac{\Bar{w}-\Bar{z}_k}{w-z_k}= \sum_k \frac{p_k^\mu p_k^\nu\varepsilon_{\mu\nu}^{+}(q)}{2p_k\cdot q} \;.
\end{equation}
Here, $q$ is the soft graviton four-momentum parametrised by (in cartesian coordinates)
\begin{equation}
    q^\mu=(1+w\bar{w},w+\bar{w},-i(w-\bar{w}),1-w\bar{w})\;,
\end{equation}
while $p_k$ is the $k$-th particle momentum 
\begin{equation}
    p^\mu=E(1+z\bar{z},z+\bar{z},-i(z-\bar{z}),1-z\bar{z})\;.
\end{equation}
Moreover, we introduced the positive helicity graviton polarization tensor given by
\begin{equation}
\varepsilon^+_{\mu\nu}=\varepsilon^+_\mu \varepsilon^+_\nu\;,\qquad \varepsilon^{+\mu}=\partial_w q^\mu\;.
\end{equation}
Plugging all this together, one can verify \eqref{eq1}.

Note that the procedure utilized here is fully general, soft theorems can be formulated for particles of any spin. Also, one could work out the subleading contributions in $\omega_g$, thus obtaining additional expressions that must hold in a scattering process, known as subleading soft theorems, here derived as Ward identities for the BMS symmetry.

\section{The Corner Proposal}\label{Sec3}

General considerations of symmetries in the previous sections gave us a top down approach to associate to any classical theory a set of charges and their associated algebras. Noether second's theorem assures that charges associated to local symmetries have support on codimension-2 surfaces, also called \textit{corners}. Therefore, it seems that the observables of the theory have to be understood and measured on these corners, on which they are defined, rather than in the entirety of spacetime. However, the main issue is that the algebra seems to depend on the problem at hand and, in particular, on the boudary conditions and gauge fixing (c.f. section \ref{bms4}). The question that thus arises is: ``\textit{Is there a universal algebra for gravity?}''. If the answer is positive, then any classical theory of gravity would have symmetries that fall in a subset of the universal ones. In section \ref{universalalgebra} we answer this question by considering embeddings of codimension-2 surfaces and in section \ref{extendedphasespace} we tackle the issue of non integrability of the charges by introducing the concept of extended phase space. For more details on the corner proposal and corners in general we refer the reader to \cite{Donnelly:2016auv, Freidel:2020xyx, Freidel:2020svx, Freidel:2020ayo, Ciambelli:2021vnn, Ciambelli:2021nmv,  Freidel:2021dxw, Ciambelli:2022cfr, Ciambelli:2022vot}.

\subsection{Universal algebra}\label{universalalgebra}
Let $M$ be a $d$-dimensional manifold and $S$ a codimension-2 manifold (the corner). The way $S$ is embedded into $M$ is defined by the injective map
\begin{equation}\label{embedding}
    \phi: S \longrightarrow M.
\end{equation}
Let $\sigma^\alpha: S \longrightarrow \mathbb{R}^{d-2},\, (\alpha = 1,...,d-2)$ be local coordinates on $S$. Let $y^M:M \longrightarrow \mathbb{R}^d$ be local coordinates on M. A choice of embedding \eqref{embedding}, is then given by $y^M(\sigma)$.
Without loss of generality (and without choosing a specific embedding) one can choose the $y^M$ coordinates such that
\begin{equation}\label{mcoordinate}
    y^M = (u^a,x^i),
\end{equation}
where $a = 1,...,d-2$ and $i=1,2$. Since we know from the covariant phase space formalism that the spacetime Lie bracket represents the symmetry algebra modulo central extensions (Equation \eqref{projectiverepresentation}), one can ask what is the maximal closed algebra by calculating the Lie bracket of two vector fields. However, since the interest lies in the codimension-2 surface, one must only calculate the bracket ``close'' to it. In order to do so, one can consider a particularly simple form of the embedding \eqref{embedding}. Once the coordinates \eqref{mcoordinate} have been chosen, this so-called ``trivial'' embedding is given by
\begin{equation}\label{trivial_embedding}
    y^M_0(\sigma) = (u_0^a(\sigma),x_0^i(\sigma)) = (0,\delta^i_\alpha \sigma^\alpha).
\end{equation}
Expanding ``close to'' this codimension-2 embedding thus corresponds to expanding around $u^a = 0$. Let us consider two vector fields $\xi, \zeta \in TM$. In the coordinate system \eqref{mcoordinate} they are expressed as
\begin{equation}
\begin{aligned}
    \xi &= \xi^a(u,x) \partial_a + \xi^i(u,x) \partial_i,\\
    \zeta &= \zeta^a(u,x) \partial_a + \zeta^i(u,x) \partial_i.
\end{aligned}
\end{equation}
 Expending close to the surface $u^a = 0$, one gets
 \begin{equation}\label{vectorsexpansion}
 \begin{aligned}
    \xi^a(u,x) &= \xi^a_{(0)}(x) + u^b\xi^a_{(1) b}(x) + u^b u^c \xi^a_{(2)b c}(x) + \mathcal{O}(u^3),\\
    \xi^i(u,x) &= \xi^i_{(0)}(x) + u^b\xi^i_{(1) b}(x) + u^b u^c \xi^i_{(2)b c}(x) + \mathcal{O}(u^3),\\
    \zeta^a(u,x) &= \zeta^a_{(0)}(x) + u^b\zeta^a_{(1) b}(x) + u^b u^c \zeta^a_{(2)b c}(x) + \mathcal{O}(u^3),\\
    \zeta^i(u,x) &= \zeta^i_{(0)}(x) + u^b\zeta^i_{(1) b}(x) + u^b u^c \zeta^i_{(2)b c}(x) + \mathcal{O}(u^3).
 \end{aligned}
 \end{equation}
Calculating their Lie bracket at second order reveals that the maximal closed sub-algebra is generated by vectors of the type
\begin{equation}\label{ucagenerators}
\begin{aligned}
    \xi &= \xi^i_{(0)} \partial_i + \qty(\xi_{(0)}^a + u^b \xi_{(1)b}^a) \partial_a,\\
    \zeta &= \zeta^i_{(0)} \partial_i + \qty(\zeta_{(0)}^a + u^b \zeta_{(1) b}^a) \partial_a.
\end{aligned}
\end{equation}
Any additional term in the expansion \eqref{vectorsexpansion} results in a non-closing algebra and is thus discarded in this analysis. The Lie bracket of the vector fields expressed in \eqref{ucagenerators} gives
\begin{equation}
\begin{aligned}
    \qty[\xi,\zeta] &= \qty[\hat{\xi}_{(0)},\hat{\zeta}_{(0)}]^i \partial_i\\
    &\qquad + \qty[\xi_{(0)}^k \zeta_{(0),k}^a - \zeta_{(0)}^k \xi_{(0),k}^a +\zeta_{(1)b}^a \xi_{(0)}^b - \xi_{(1)b}^a \zeta_{(0)}^b] \partial_a\\
    &\qquad + u^b\qty[-\qty[\xi_{(1)},\zeta_{(1)}]^a{}_{\, b} + \xi_{(0)}^k \zeta_{(1)b,k}^a - \zeta_{(0)}^k \xi_{(1)b,k}^a] \partial_a, 
\end{aligned}
\end{equation}
where $\hat{\zeta} \equiv \zeta^j\partial_j$ and the comma denotes partial differentiation. Remembering that the indices $i,j,k,...$ denote coordinates on the surface and $a,b,c,...$ denote coordinates on $M$ in the directions normal to the surface, it is now easy to interpret this result. The first line simply corresponds to diffeomorphisms on the surface. The first part of the second line corresponds to surface diffeomorphisms acting on the normal translations and the second part to the general linear group acting on normal translations. Finally, the first part of the last line corresponds simply to the general linear algebra, and the second part to the surface diffeomorphisms acting on this general linear algebra. Thus the complete group corresponding to this maximally closed algebra is
\begin{equation}\label{universalgroup}
    \mathrm{UCS} = \qty(\mathrm{Diff}(S) \ltimes  \mathrm{GL}(2,\mathbb{R})^{S})\ltimes \qty(\mathbb{R}^2)^S.
\end{equation}
UCS stands for the \textit{Universal Corner Symmetries}. A few remarks on this result are in order:
\begin{itemize}
    \item The semi-direct product emphasizes that, while the corner diffeomorphisms act on the linear transformations and the translations, the converse is not true. Similarly for the linear group acting on the translations.
    \item The $S$ in the exponent of the linear group and the translations means that there is a copy of the general linear group and the translation group at each point of the corner. This is the origin of supertranslations in section \ref{charges}.
    \item One can see that the BMS and BMSW groups discussed in section \ref{bms4} are a subgroup of the universal one. Note that the two copies of supertranslations in the UCS arise from the fact that the corner is a codimension-2 surface and we therefore have two normal directions. However, in the case of asymptotic infinity, one can only move along the $\mathfrak{I}$ boundary and we are thus left with only one copy of  supertranslations.
    \item This entire analysis was done without the need of introducing a metric on the manifold. Mathematically this results is topological and not geometrical, which translates to kinematical and not dynamical in the physical realm. Therefore this algebra does not depend on the particular theory of gravity at hand and is thus called \textit{universal}.
\end{itemize}
It is believed in the corner community that this algebra is the most general one that can have charges \cite{Ciambelli:2021vnn, Ciambelli:2022cfr}, providing a bottom up approach to quantum gravity as a condition that any theory should hold a representation of \eqref{universalgroup} at some level. For example, the connection of this symmetry group to Loop Quantum Gravity is discussed in \cite{Freidel:2023bnj}.
\subsection{Extended phase space} \label{extendedphasespace}
\subsubsection{Keeping track of the embedding}
Let us briefly recap the results up to this point. We know that the UCS Lie bracket realises the algebra of the charges modulo central extensions. The next step towards the quantization of the theory would be to promote the charges to hermitian operators on some Hilbert space and replace the algebra brackets by Dirac brackets. However there is another issue here. In general, as we have seen in previous sections, there might be fluxes present. Mathematically, this translates to the non-closure of the algebra brackets, and thus we do not know how to quantize the theory. This issue arises from a careless treatment of the embedding in the action. More precisely, we assumed something of the following form in a number of computations:
\begin{equation}\label{variationandintegrationcommute}
    \fL_{\hat\xi} \int_S \mathcal{A} = \int_S \fL_{\hat\xi} \mathcal{A},
\end{equation}
for a general functional $\mathcal{A}$.
However, the commutation of the variation and the integral is only true in the case where the variation does not move the surface we are integrating on. A way of being more careful about it is to keep track of the embedding \eqref{embedding} when integrating a quantity on the surface:

\begin{equation}
\fL_{\hat\xi} \int_S \phi^*\qty(\mathcal{A}) = \int_S \phi^* \qty(\fL_{\hat\xi} \mathcal{A}) + \int_S \qty(\fL_{\hat\xi} \phi^*) \qty(\mathcal{A}),
\end{equation}
where we see the appearance of a second term that was not present in equation \eqref{variationandintegrationcommute}. The question now becomes: ``\textit{How can we keep track of the variation of the embedding at the phase space level?}'' This is done by the introduction of a new field $\chi$ which is a vector in spacetime and a one-form on field space:
\begin{equation}
    \chi \in TM \otimes T^*\Gamma,
\end{equation}
defined through the relation
\begin{equation}\label{chicondition}
    I_{\hat{\xi}} \chi = -\xi,
\end{equation}
This field keeps track of the variation of the embedding in the following sense: For any functional $\mathcal{F}$ on field space, we have
\begin{equation} \label{embeddingvariation}
    \delta \phi^* (\mathcal{F}) = \phi^* (\ldv{\chi} \mathcal{F}).
\end{equation}
Note that this relation can be thought off as the equivalence between the passive and the active interpretation of diffeomorphisms. The $\chi$ field, which is sometimes called an \textit{edge mode} because it lives on the corner, is then added to the phase space of the theory.

\subsubsection{Extended symplectic form} 
We will now calculate again the symplectic form but on the extended phase space that includes the new field. In the following derivation we will use the following relations that are proved in appendix \ref{proofofimportantrelations}:
\begin{align}
    \delta \chi &= -\frac{1}{2} [\chi,\chi], \label{equationa}\\
    \ldv{\chi} \iota_\chi L &= \frac{1}{2} \iota_{\qty[\chi,\chi]}L + \frac{1}{2} \dd (\iota_\chi i _\chi L), \label{equationb}\\
    I_{\hat{\xi}} \iota_{\delta \chi} L &= \iota_{\qty[\xi,\chi]} L.\label{equationc}
\end{align}
Consider the action defined in a subregion $R$ that allows us to keep track of the embedding:
\begin{equation}
    S_R = \intpull{R}{d} \qty(L)\;,
\end{equation}
where $L$ is a functional of the dynamical fields $\varphi$ and their derivatives.
Taking its variation yields
\begin{equation}
    \begin{aligned}
        \delta S_R &= \int_R \qty[\delta \phi^* (L) + \phi^* (\delta L)]\\
        &= \intpull{R}{d}(\ldv{\chi} L +\delta L)\\
        &\hat{=} \intpull{R}{d}(\dd \iota_\chi L + \dd \theta) \\
        &= \intpull{R}{d} (\dd \theta^{\mathrm{ext}})\;,
    \end{aligned}
\end{equation}
where one should stress that the extended symplectic potential is now a function of the dynamical fields $\varphi$, their variation $\delta\varphi$,  as well as the new field $\chi$, that is, $\theta^{\mathrm{ext}}[\varphi, \delta \varphi, \chi]$.
The extended symplectic potential is now given by
\begin{equation}
    \Theta^{\mathrm{ext}}= \intpull{\Sigma}{d-1} \theta^{\mathrm{ext}} = \intpull{\Sigma}{d-1} (\iota_\chi L + \theta).
\end{equation}
The extended symplectic two-form can be calculated:
\begin{equation}
    \begin{aligned}
        \Omega^{\mathrm{ext}}&= \delta \Theta^{\mathrm{ext}}\\
        &= \int_\Sigma [\delta \phi^*(\iota_\chi L + \theta) + \phi^*_(\delta(\iota_\chi L+ \theta))]\\
        &= \intpull{\Sigma}{d-1} (\ldv{\chi} \iota_\chi L + \ldv{\chi} \theta + \iota_{\delta \chi} L - \iota_\chi \delta L + \delta \theta)\\
        & \he \intpull{\Sigma}{d-1}(\ldv{\chi} \iota_\chi L + \dd \iota_\chi \theta + \delta \theta + \iota_{\delta \chi} L) \\
        &= \intpull{\Sigma}{d-1} (\delta \theta + \frac{1}{2}\dd (\iota_\chi \iota_\chi L) + \dd \iota_\chi \theta)\\
        &= \intpull{\Sigma}{d-1} \left(\delta \theta + \dd (\iota_\chi \theta + \frac{1}{2} \iota_\chi \iota_\chi L)\right),
    \end{aligned}
\end{equation}
where we have used equations \eqref{equationa} and \eqref{equationb} to go from the fourth to the fifth line.
One can thus rewrite the extended symplectic form in the following way:
\begin{equation}
    \Omega^{\mathrm{ext}}= \Omega + \intpull{S}{d-2} (\iota_\chi \theta + \frac{1}{2} \iota_\chi \iota_\chi L).
\end{equation}
The additional term is thus a corner contribution, which will render all charges integrable. Loosely speaking, the additional field $\chi$ renders the flux part of the augmented phase space. Indeed,
let us now calculate the contraction \eqref{hamiltonian vector field} within this extended phase space formalism:
\begin{equation}
    \begin{aligned}
        I_{\hat{\xi}} \Omega^{\mathrm{ext}} &= I_{\hat{\xi}} \intpull{\Sigma}{d-1} (\delta (\theta + \iota_\chi L) + \ldv{\chi} (\theta + \iota_\chi L))\\
        &= \intpull{\Sigma}{d-1}(I_{\hat{\xi}} \delta \theta + I_{\hat{\xi}} \ldv{\chi} \theta + I_{\hat{\xi}} \ldv{\chi} \iota_\chi L + I_{\hat{\xi}} \iota_{\delta \chi} L - I_{\hat{\xi}} \iota_\chi \delta L)\\
        &= \intpull{\Sigma}{d-1}(\ldv{\xi} \theta - \delta I_{\hat{\xi}} \theta + I_{\hat{\xi}} \ldv{\chi} \theta + I_{\hat{\xi}}\ldv{\chi}\iota_\chi L + \iota_{\qty[\xi,\chi]}L - I_{\hat{\xi}} \iota_\chi \delta L)\\
        &= \intpull{\Sigma}{d-1}(- \delta I_{\hat{\xi}} \theta - \ldv{\chi} I_{\hat{\xi}} \theta + I_{\hat{\xi}} \ldv{\chi} \iota_\chi L + \iota_{\qty[\xi,\chi]}L - I_{\hat{\xi}} \iota_\chi \delta L)\\
        &= \intpull{\Sigma}{d-1} (- \delta I_{\hat{\xi}} \theta - \ldv{\chi} I_{\hat{\xi}} \theta - \ldv{\xi} \iota_\chi L + \ldv{\chi} \iota_{\xi} L + \iota_{\qty[\xi,\chi]}L + \iota_{\xi} \delta L + \iota_\chi I_{\hat{\xi}} \delta L)\\
        &= \intpull{\Sigma}{d-1}(-\delta I_{\hat{\xi}} \theta - \ldv{\chi} I_{\hat{\xi}} \theta + \ldv{\chi} \iota_{\xi} L + \iota_{\xi} \delta L)\\
        &= \intpull{\Sigma}{d-1}( - \delta I_{\hat{\xi}} \theta - \ldv{\chi} I_{\hat{\xi}}\theta + \ldv{\chi} \iota_{\xi} L + \iota_{\xi} \delta L)\\
        &= - \intpull{\Sigma}{d-1}(\delta (I_{\hat{\xi}}\theta - \iota_{\xi} L) + \ldv{\chi} (I_{\hat{\xi}} - \iota_{\xi} L)),
        \end{aligned}
\end{equation}
where in the second line we have used \eqref{equationc} and in the third line we used that the spacetime Lie derivative coincide with the field-space Lie derivative for $L$ and $\theta$. Recalling the expression for the Noether current associated to a diffeomorphism:
\begin{equation}
    J_{\hat{\xi}} = I_{\hat{\xi}} \theta - \iota_{\xi} L,
\end{equation}
we can rewrite the contraction as
\begin{equation}
    \begin{aligned}
        I_{\hat{\xi}} \Omega^{\mathrm{ext}} &= - \intpull{\Sigma}{d-1} (\delta J_{\hat{\xi}} + \ldv{\chi} J_{\hat{\xi}})\\
        &= - \delta (\intpull{\Sigma}{d-1} (J_{\hat{\xi}}))\\
        &= -\delta H_{\hat\xi}.
    \end{aligned}
\end{equation}
where we have used \eqref{embeddingvariation} one last time and the charge $H_{\hat\xi}$ is the same as in the non-extended phase space. We thus see that introducing this new field takes care of the term that  spoils integrability. This crucial result established in \cite{Ciambelli:2021nmv} implies that the charge algebra is represented by the standard Poisson bracket, see \cite{Freidel:2021dxw}. This is a remarkable feature, because the system is still dissipative, albeit integrable. Rendering all diffeomosphisms integrable is a turning point in the corner proposal because it prepares the classical gravitational setup to finally address quantum gravity, as we will discuss below.

\subsection{The corner proposal for quantum gravity}\label{corner proposal sec}

We here enunciate the main idea of the corner proposal, and offer some conclusive remarks, to summarize the journey taken.

We are finally ready to state the proposal. It is hard to formulate a consistent theory of quantum gravity, and therefore it would be very useful to have a bottom-up approach to answer some questions in this field.
The main idea of the corner proposal is the following:

\begin{center}
    \emph{Gravity is described by a set of charges and their algebra at corners}. 
\end{center}

\noindent Here, corners refer to any codimension-2 surface. This proposal instructs us to loose ties with classical concepts and notions, that would not survive in the quantum realm. For example, a full bulk metric is at best a semi-classical notion. What we have shown to be more robust is the concept of corners, on which charges are defined. Even more fundamental is the concept of symmetries and algebras. In this regard, it was instrumental to enlarge the phase space, such that all diffeomorphisms are canonically realized without fluxes. There, the  Universal Corner Symmetry algebra can be derived, and the corner proposal posits that it survives in the quantum regime and becomes the algebra of observables. We must therefore focus on charges and their Poisson brackets, which are then promoted to operators in the quantum theory, with the usual prescription $\poissonbracket{Q_{\hat\xi}}{Q_{\hat\zeta}}\rightarrow \frac{1}{i\hbar}[Q_{\hat\xi},Q_{\hat\zeta}]$. We can then study the representation theory of the charge algebra, and investigate whether there are unitary representations that could be suitable to describe the quantum geometry.
The classical spacetime emerges from one particular representation of this algebra, but there are other representations, that are relevant for the quantum theory. Obviously, this is not an easy task, but it is a promising and concrete program toward a better understanding of quantum gravity.

From the algebraic perspective, the mission of the corner proposal is clear. Starting from the Lie bracket of two symmetry generators, we have a way, using the covariant phase space, to descend to the Poisson brackets of the system. This is already a delicate and mathematically challenging step. The endpoint is a projective representation of the symmetry algebra, at the level of charges. From there, one can pursue a canonical quantization procedure to derive the Dirac bracket. Which algebra should we focus on? If each quantization procedure leads to a different algebra, than this proposal stops being predictive. Thankfully, we have shown that the Universal Corner Symmetry algebra is very robust and maximal, which therefore means that it is independent of a particular setup and classical Lagrangian: the latter can only further reduce the charged symmetries. Therefore, we have a narrow window to probe quantum gravity: study the Universal Corner Symmetry algebra in particular setups. Clearly, loosing the concept of metric and manifold opens the door to the idea of quantum geometries, and hopefully this could represent a new avenue of investigation for the quantum gravity community. 

Let us quickly summarize the results exposed. We have shown that the covariant phase space is the suitable arena to study Noether's theorems. The nature of a local symmetry is encoded in its corner charge. If the latter is non-zero, then this is a physical symmetry. Gravity makes the study of the nature of the symmetry hard, because it brings complications, such as integrability issues, divergences, and dissipation. At the end, after curing those, we have shown how symmetries are the crucial ingredients from which one can derive physical effects, such as memory effects and soft theorems. In this regard, the infrared triangle should rather be depicted as a symmetries pyramid, where the concept of symmetries lies on the top of the pyramid, and all the rest can be derived from it (see \cref{fig:Pyr}). Using this as a guiding principle, we arrived to the corner proposal, which is the idea that symmetries are the ultimate clues toward a bottom-up understanding of quantum gravity. We exposed these ideas in these lecture notes, and in particular we discussed how to prepare the classical milieu using only tools that are expected to survive in quantum gravity, such as algebras, charges, and symmetries. In this respect, the derivation of the Universal Corner Symmetry algebra and the extension of the phase space to make all diffeomorphisms integrable are important steps, to close the classical gravity chapter and open the quantum gravity one.

\vspace{2em}

\begin{figure}[hbt!]
\begin{center}
\includegraphics[scale=0.4]{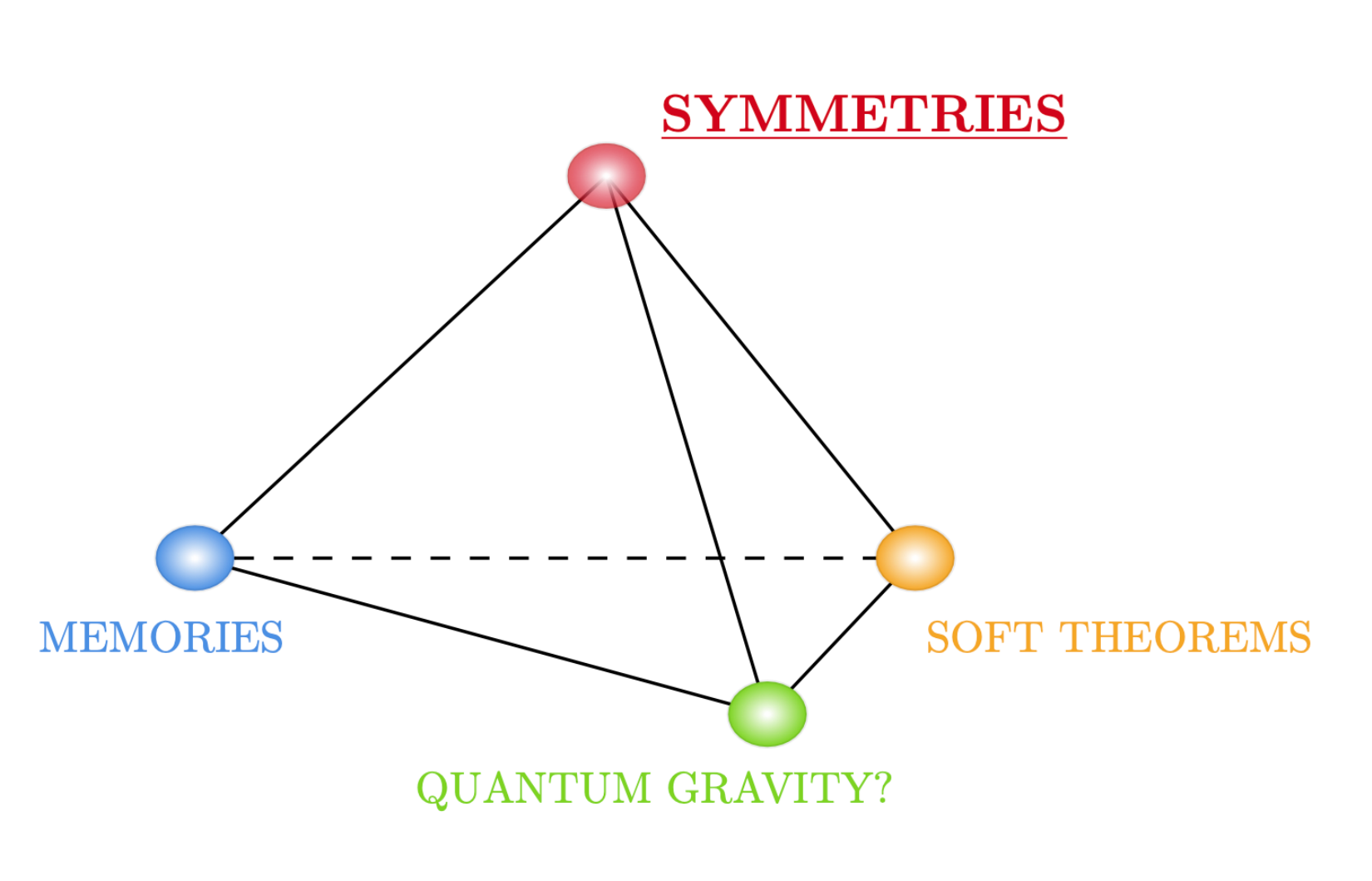}\caption{Rather than an infrared triangle, our approach culminates in ``the symmetries pyramid''. Indeed, the Corner Proposal puts emphasis on the symmetries of the theory, from which everything can be derived.}\label{fig:Pyr}
\end{center}
\end{figure}

\paragraph{Acknowledgements} We would like to thank the organisers and the participants of the 59 Winter School of Theoretical Physics and third COST Action CA18108 Training School for the very stimulating atmosphere. Research at Perimeter Institute is supported in part by the Government of Canada through the Department of Innovation, Science and Economic Development Canada and by the Province of Ontario through the Ministry of Colleges and Universities. This work has been partially supported by the Regional Government of Castilla y León (Junta de Castilla y León), by the Ministry of Science and Innovation MICIN and the European Union NextGenerationEU (PRTR C17.I1) and by the Faculty of Physics, University of Belgrade (grant number 451-03-47/2023-01/200162).

\appendix
\section{Proof of Important Relations}\label{proofofimportantrelations}
The first relation we will look at is the variation of the field $\chi$. Consider a general integrated functional:
\begin{equation}
    \mathcal{A} = \intpull{R}{d} (\mathcal{F}),
\end{equation}
where $R$ denotes a subregion of spacetime which makes it simpler to keep track of the embedding. We have
\begin{equation}
\begin{aligned}
    \delta \mathcal{A} &= \int_R [\delta \phi^*(\mathcal{F}) + \phi^*(\delta\mathcal{F})]\\
    &= \intpull{R}{d} (\delta \mathcal{F} + \ldv{\chi}\mathcal{F}).
\end{aligned}
\end{equation}
Taking the variation a second time yields
\begin{equation}
    \begin{aligned}
        0 = \delta^2 \mathcal{A} &= \int_R [\delta \phi^*(\delta \mathcal{F} + \ldv{\chi} \mathcal{F}) + \phi^* (\delta (\ldv{\chi}\mathcal{F}))]\\
        &= \intpull{R}{d} [\ldv{\chi} \delta \mathcal{F} + \ldv{\chi}\ldv{\chi} \mathcal{F} + \ldv{\delta \chi} \mathcal{F}- \ldv{\chi} \delta \mathcal{F}]\\
        &= \intpull{R}{d} [(\ldv{\chi} \ldv{\chi} + \ldv{\delta \chi}) \mathcal{F}] \\
        &= \intpull{R}{d} [(\ldv{\frac{1}{2}[\chi,\chi]}  + \ldv{\delta \chi}) \mathcal{F}].
    \end{aligned}
\end{equation}
Note that the minus sign in front of the last term of the second line comes from the fact that $\chi$ is a field space one-form. We thus have
\begin{equation}\label{chivariation}
    \delta \chi = -\frac{1}{2}[\chi,\chi].
\end{equation}
The second relation is the following
\begin{equation}
    \ldv{\chi} \iota_\chi L = \frac{1}{2} \iota_{\qty[\chi,\chi]}L + \frac{1}{2} \dd (\iota_\chi \iota _\chi L),
\end{equation}
which comes from the identity $[\ldv{\xi_1},i_{\xi_2}] = i_{[\xi_1,\xi_2]}$:
\begin{equation}
\begin{aligned}
    \ldv{\chi} \iota_\chi L &= [\ldv{\chi},\iota_\chi] L - \iota_\chi \ldv{\chi} L \\
    &= \iota_{\qty[\chi,\chi]} L - \iota_\chi \dd \iota_\chi L\\
    &= \iota_{\qty[\chi,\chi]} L - \ldv{\chi}(\iota_\chi L) + \dd (\iota_\chi \iota_\chi L).
\end{aligned}
\end{equation}
The third idendity that we will use is
\begin{equation}
\begin{aligned}
    I_{\hat{\xi}} \iota_{\delta \chi} L = -\frac{1}{2} I_{\hat{\xi}} \iota_{\qty[\chi,\chi]} L = \iota_{\qty[\xi,\chi]} L,
\end{aligned}
\end{equation}
where \eqref{chicondition} and \eqref{chivariation} were used. 

\providecommand{\href}[2]{#2}\begingroup\raggedright\endgroup

\end{document}